\documentclass[11pt]{article}
\usepackage{amsmath,amsfonts,amssymb,graphicx,epsfig,pdflscape}
\usepackage[usenames]{color}
\usepackage{pstricks}
\usepackage{arydshln,cite}
\numberwithin{equation}{section}

\newcommand{\nc}{\newcommand}
\nc\disp{\displaystyle}
\nc{\fh}{\hat{f}}
\nc{\muh}{\hat{\mu}}
\nc{\nuh}{\hat{\nu}}
\nc{\spos}[2]{\makebox(0,0)[#1]{$\sm{#2}$}}
\nc{\sm}[1]{{\scriptstyle #1}}
\nc{\bib}{\bibitem}
\nc{\al}{\alpha}
\nc{\g}{\gamma}
\nc{\G}{\Gamma}
\nc{\D}{\Delta}
\nc{\eps}{\epsilon}
\nc{\la}{\lambda}
\nc{\La}{\Lambda}
\nc{\var}{\varphi}
\nc{\pa}{\partial}
\nc{\nn}{\nonumber \\ }
\nc{\hf}{\frac{1}{2}}
\nc{\dz}{\frac{dz}{2\pi i}}
\nc{\bin}[2]{\left(\!\!\!\begin{array}{c} {#1}\\ {#2} \end{array}\!\!\!\right)}
\nc{\be}{\begin{equation}}
\nc{\ee}{\end{equation}}
\nc{\bea}{\begin{eqnarray}}
\nc{\eea}{\end{eqnarray}}
\nc{\bra}[1]{\langle {#1}|}
\nc{\ket}[1]{|{#1}\rangle}
\nc{\ketw}[1]{({#1})^{\phantom{a}}_{{\cal W}}}
\nc{\ketwa}[1]{({#1})^{\ast}_{{\cal W}}}
\nc{\chit}{\raisebox{0.25ex}{$\chi$}}
\nc{\chih}{\raisebox{0.25ex}{$\hat\chi$}}
\nc{\Dti}{\tilde{\D}}
\nc{\Db}{\mbox{\boldmath $D$}}
\nc{\Hb}{\mbox{\boldmath $H$}}
\nc{\Ib}{\mbox{\boldmath $I$}}
\nc{\qb}{\bar{q}}
\nc{\Ac}{\mathcal{A}}
\nc{\Bc}{\mathcal{B}}
\nc{\Cc}{\mathcal{C}}
\nc{\Dc}{\mathcal{D}}
\nc{\Ec}{\mathcal{E}}
\nc{\Fc}{\mathcal{F}}
\nc{\Ic}{\mathcal{I}}
\nc{\Jc}{\mathcal{J}}
\nc{\Mc}{\mathcal{M}}
\nc{\Oc}{\mathcal{O}}
\nc{\Pc}{\mathcal{P}}
\nc{\Rc}{\mathcal{R}}
\nc{\Vc}{\mathcal{V}}
\nc{\Wc}{\mathcal{W}}
\nc{\Xc}{\mathcal{X}}
\nc{\Yc}{\mathcal{Y}}
\nc{\Zc}{\mathcal{Z}}
\nc{\fus}{\mbox{}\,\hat\otimes\,\mbox{}}
\nc{\Rh}{\hat{\mathcal{R}}}
\nc{\Mh}{\hat{M}}
\nc{\Qh}{\hat{Q}}
\nc{\Cch}{\hat{\Cc}}
\def\vvdots{\mathinner{\mkern1mu\raise1pt\vbox{\kern7pt\hbox{.}}\mkern2mu
  \raise4pt\hbox{.}\mkern2mu\raise7pt\hbox{.}\mkern1mu}}
\nc{\gauss}[2]{\left[\!\!\begin{array}{c} {#1}\\ {#2} \end{array}\!\!\right]}
\nc{\sbin}[2]{\left\{\!\!\!\begin{array}{c} {#1}\\ {#2} 
\end{array}\!\!\!\right\}}
\nc{\sbinlr}[2]{\Big\langle\!\!\begin{array}{c} {#1}\\ {#2} 
\end{array}\!\!\Big\rangle}
\nc{\bino}[2]{\left(\!\!\begin{array}{c} {#1}\\ {#2} \end{array}\!\!\right)}
\def\half {\mbox{$\textstyle \frac{1}{2}$}}

\nc{\ch}{{\rm ch}}
\nc{\R}{{\cal R}}
\nc{\dkk}{\delta_{j,\{k,k'\}}^{(2)}}
\nc{\drr}{\delta_{j,\{r,r'\}}^{(2)}}
\nc{\ddkk}{\delta_{j,\{k,k'\}}^{(4)}}
\nc{\dddkk}{\delta_{j,\{k,k'\}}^{(8)}}
\nc{\dnn}{\delta_{j,\{n,n'\}}^{(2)}}
\nc{\ddnn}{\delta_{j,\{n,n'\}}^{(4)}}
\nc{\dddnn}{\delta_{j,\{n,n'\}}^{(8)}}
\definecolor{lightlightblue}{rgb}{.85,.85,1}
\definecolor{midblue}{rgb}{.7,.7,1}

\begin{document}

\topmargin -5mm
\oddsidemargin 5mm

\setcounter{page}{1}

\vspace{8mm}
\begin{center}
{\Large {\bf $\Wc$-extended Kac representations and integrable boundary conditions}}
\\[.3cm]
{\Large {\bf in the logarithmic minimal models ${\cal WLM}(1,p)$}}

\vspace{10mm}
{\LARGE J{\o}rgen Rasmussen}
\\[.3cm]
{\em Department of Mathematics and Statistics, University of Melbourne}\\
{\em Parkville, Victoria 3010, Australia}
\\[.4cm]
{\tt j.rasmussen\,@\,ms.unimelb.edu.au}
\end{center}

\vspace{8mm}
\centerline{{\bf{Abstract}}}
\vskip.4cm
\noindent
We construct new Yang-Baxter integrable boundary conditions in the lattice approach to
the logarithmic minimal model ${\cal WLM}(1,p)$ giving rise to reducible
yet indecomposable representations of rank 1 in the continuum scaling limit. 
We interpret these $\Wc$-extended Kac representations as finitely-generated
$\Wc$-extended Feigin-Fuchs modules over the triplet $W$-algebra $\Wc(p)$.
The $\Wc$-extended fusion rules of these representations
are inferred from the recently conjectured Virasoro fusion rules of the Kac representations
in the underlying logarithmic minimal model ${\cal LM}(1,p)$.
We also introduce the modules contragredient to the $\Wc$-extended Kac modules
and work out the correspondingly-extended fusion algebra.
Our results are in accordance with the Kazhdan-Lusztig dual of tensor products of modules
over the restricted quantum universal enveloping algebra $\bar{U}_q(sl_2)$ at $q=e^{\pi i/p}$.
Finally, polynomial fusion rings isomorphic with the various fusion algebras are determined,
and the corresponding Grothendieck ring of characters is identified.

\renewcommand{\thefootnote}{\arabic{footnote}}
\setcounter{footnote}{0}

\newpage
\tableofcontents

\section{Introduction}

Physical systems described by logarithmic conformal field theories
(CFTs)~\cite{Gur9303,Flohr0111,Gab0111} include 
polymers~\cite{Saleur87a,Duplantier86,Saleur92,PR0610,Nigro0903,PRV0910},
percolation~\cite{Cardy92,Watts9603,Cardy0103,Smirnov01,MatRidout0708,Ridout0808,DubJacSal1001},
symplectic fermions~\cite{Kausch9510,Kausch0003}
and the abelian sandpile 
model~\cite{Dhar99,MahRuelle0107,Ruelle0203,MRR0410,PGPR0710}.
In fact, an infinite series of logarithmic CFTs arises in the continuum scaling limit of certain
two-dimensional lattice models of non-local statistical mechanical systems
at criticality~\cite{PRZ0607}. 
Polymers~\cite{PR0610,PRV0910}, 
percolation~\cite{RP0706,RP0804} and symplectic fermions~\cite{PRR0803}
are all described by these logarithmic minimal models.
Quantum spin chains with a non-diagonalizable 
Hamiltonian~\cite{RS0701} likewise give rise to logarithmic CFTs. 

Mathematically, vertex operator algebras (VOAs)~\cite{Bor86,FLM88,Kac01,FB-Z01}
provide an algebraic pendant to CFTs. The abelian category of modules over
a VOA associated with a rational CFT is semi-simple and contains
only finitely many simple objects (irreducible representations).
In order for a CFT to make sense on a higher-genus Riemann surface,
the corresponding VOA must satisfy Zhu's $C_2$-cofiniteness condition~\cite{Zhu96}.
The triplet $W$-algebra $\Wc(p)$~\cite{Kau91}, where $p=2,3,\ldots$, is an example of 
such a VOA, as demonstrated in~\cite{CF0508,AM0707}.
The corresponding abelian category of modules is {\em non}-semi-simple, however,
and the associated CFT is 
{\em logarithmic}~\cite{Kausch9510,GK9606,FHST0306,GR0707}.
VOAs of logarithmic CFTs are discussed more generally 
in~\cite{Milas0101,Miy0209,HLZ0710,Huang0712,Huang09}.

In a series of papers~\cite{FGST0504,FGST0512,FGST0606hep,FGST0606math}, 
Feigin et al conjectured and examined a Kazhdan-Lusztig duality
between the logarithmic CFTs based on the triplet $W$-algebra $\Wc(p)$
and the representation theory of the restricted quantum universal enveloping algebra 
$\bar{U}_q(sl_2)$ at $q=e^{\pi i/p}$~\cite{Sut94,CP94,Xiao97,Arike0706}.
It was subsequently proven~\cite{NT0902} that the category of $\Wc(p)$-modules and
the category of finite-dimensional $\bar{U}_q(sl_2)$-modules indeed are equivalent
as abelian categories for all $p=2,3,\ldots$.
For $p>2$, they are not, however, equivalent as braided quasi-tensor categories since
their natural tensor structures are not fully compatible~\cite{KS0901}.

The present work concerns the logarithmic minimal models ${\cal LM}(p,p')$~\cite{PRZ0607} 
in the $\Wc$-extended picture~\cite{PRR0803,RP0804,Ras0805} where they are
denoted by ${\cal WLM}(p,p')$. The parameters $p$ and $p'$ constitute 
a pair of coprime positive integers $1\leq p<p'$, and focus here is on the case ${\cal WLM}(1,p')$.
For simplicity, this is denoted by ${\cal WLM}(1,p)$ where $p=2,3,\ldots$, and
the extension is believed to be with respect to the triplet $W$-algebra $\Wc(p)$.
The logarithmic minimal model ${\cal WLM}(1,p)$ is thus conjectured to be associated
with the VOA based on $\Wc(p)$, and we will assert this in the following.

Associated with a Yang-Baxter integrable boundary condition
in the lattice approach to ${\cal LM}(1,p)$, there is a so-called Kac representation $(r,s)$ for each
pair of positive Kac labels $r,s\in\mathbb{N}$.
The corresponding Virasoro modules were identified in~\cite{Ras1012}
and conjectured to be finitely-generated Feigin-Fuchs modules~\cite{FF89}.
The fusion algebras generated by the Kac representations and their
contragredient counterparts $(r,s)^\ast$ were also determined in~\cite{Ras1012}
and confirmed in~\cite{BGT1102} based on the Kazhdan-Lusztig duality
conjectured in~\cite{BFGT0901}.
Here we lift the findings of~\cite{Ras1012} to the $\Wc$-extended picture using
methods developed in~\cite{PRR0803}.

In Section~\ref{SecLM1p}, we review the logarithmic minimal models
${\cal LM}(1,p)$ in the Virasoro picture. Following~\cite{Ras1012},
we discuss the fusion properties of the Kac representations and their
contragredient counterparts.
In Section~\ref{SecW}, we generalize the construction of $\Wc$-extended
representations in~\cite{PRR0803} from $\Wc$-irreducible to general 
$\Wc$-extended Kac representations $\ketw{r,s}$ where $r,s\in\mathbb{N}$. 
We thus introduce new Yang-Baxter integrable boundary conditions
whose continuum scaling limits give rise to these $\Wc$-extended Kac representations. 
We interpret these representations as finitely-generated
$\Wc$-extended Feigin-Fuchs modules over the triplet $W$-algebra $\Wc(p)$.
The contragredient modules $\ketwa{r,s}$ of the $\Wc$-extended Kac
representations $\ketw{r,s}$ are also introduced, and the various fusion rules are
inferred from the recently conjectured Kac fusion algebra in the Virasoro picture~\cite{Ras1012}.
This Kac fusion algebra is summarized in Appendix~\ref{AppFusVir}.
In Section~\ref{SecPol}, we determine polynomial fusion rings isomorphic with the 
$\Wc$-extended Kac fusion algebra and its contragredient extension.
We also identify the corresponding Grothendieck ring associated with the
Virasoro characters of the $\Wc$-extended representations.
Section~\ref{SecDisc} contains some concluding remarks and a comparison
of our results on fusion with the tensor structure of the restricted quantum universal 
enveloping algebra $\bar{U}_q(sl_2)$ at $q=e^{\pi i/p}$~\cite{KS0901}.
To facilitate this comparison, a dictionary relating the different notations
is presented in Appendix~\ref{AppDic}.

\subsection*{Notation}

For $n,m\in\mathbb{Z}$ and modules $A_n$,
\be
\begin{array}{rcl}
 &\mathbb{Z}_{n,m}=\mathbb{Z}\cap[n,m],\qquad \mathbb{N}_0=\mathbb{N}\cup\{0\}&
\\[.3cm]
 &\displaystyle{\eps(n)=\frac{1-(-1)^n}{2},\qquad  n\cdot m=1+\eps(n+m)},\qquad
  \displaystyle{\bigoplus_n^N A_n=\bigoplus_{n=\eps(N),\,\mathrm{by}\,2}^N A_n}&
\end{array}
\ee
It is noted that $n\cdot m\in\mathbb{Z}_{1,2}$
and that this dot product is commutative and associative.

\section{Logarithmic minimal model ${\cal LM}(1,p)$}
\label{SecLM1p}

The logarithmic minimal model ${\cal LM}(1,p)$ is a logarithmic CFT
with central charge
\be
 c=1-6\frac{(p-1)^2}{p},\qquad p=2,3,\ldots
\label{c}
\ee
In this section, we review the Virasoro representations 
associated with the boundary conditions appearing in the lattice approach to  
${\cal LM}(1,p)$ as described in~\cite{PRZ0607,RP0707,Ras1012}.
We also recall the associated contragredient modules introduced in~\cite{Ras1012} and
review the corresponding fusion algebras.

\subsection{Kac representations}
\label{SecKacRep}

There is a so-called Kac representation $(r,s)$ for each pair of positive Kac labels
$r,s\in\mathbb{N}$. It is associated with a Yang-Baxter integrable boundary condition in the lattice
approach to ${\cal LM}(1,p)$~\cite{PRZ0607,RP0707} and arises in the continuum
scaling limit. A classification of these Kac representations as modules over the Virasoro algebra
was recently proposed in~\cite{Ras1012}. It was thus conjectured that they can be viewed as
finitely-generated submodules of Feigin-Fuchs modules~\cite{FF89}.

To describe these finitely-generated Feigin-Fuchs modules, we first consider the quotient module
\be
 Q_{r,s}=V_{r,s}/V_{r,-s},\qquad r,s\in\mathbb{N}
\ee
where $V_{r,s}$ is the Verma module of conformal weight
\be
 \D_{r,s}=\frac{(rp-s)^2-(p-1)^2}{4p},\qquad r,s\in\mathbb{Z}
\ee
The corresponding {\em irreducible} highest-weight module is denoted by $M_{r,s}$,
where we set $M_{r,0}=M_{0,s}=0$.
Parameterizing the second Kac label as
\be
 s=s_0+kp,\qquad s_0\in\mathbb{Z}_{1,p-1},\quad k\in\mathbb{N}_0
\label{s}
\ee
the structure diagram of the quotient module $Q_{r,s}$ is given by
\be
 Q_{r,s}:\quad\ \
  M_{k-r+1,p-s_0}\to M_{k-r+2,s_0}\to M_{k-r+3,p-s_0}\to\ldots\to M_{k+r-1,p-s_0}\to M_{k+r,s_0}
\ee

We can associate a pair of finitely-generated
Feigin-Fuchs modules to every quotient module $Q_{r,s}$.
For $2r-1<2k$, the Feigin-Fuchs modules corresponding to $Q_{r,s}$ are characterized by the
structure diagrams
\be
\begin{array}{rcl}
 &Q^{\to}_{r,s}:&\quad
  M_{k-r+1,p-s_0}\to M_{k-r+2,s_0}\gets M_{k-r+3,p-s_0}\to\ldots\gets M_{k+r-1,p-s_0}\to 
   M_{k+r,s_0}
 \\[.5cm]
 &Q^{\gets}_{r,s}:&\quad
  M_{k-r+1,p-s_0}\gets M_{k-r+2,s_0}\to M_{k-r+3,p-s_0}\gets\ldots\to M_{k+r-1,p-s_0}\gets 
    M_{k+r,s_0}
\end{array}
\label{2r<2k}
\ee
For $2r-1>2k$, the Feigin-Fuchs modules corresponding to $Q_{r,s}$ are characterized by the
structure diagrams
\be
\begin{array}{rcl}
 &Q^{\to}_{r,s}:&\quad
  M_{r-k,s_0}\to M_{r-k+1,p-s_0}\gets M_{r-k+2,s_0}\to\ldots\to M_{r+k-1,p-s_0}\gets M_{r+k,s_0}
 \\[.5cm]
 &Q^{\gets}_{r,s}:&\quad
  M_{r-k,s_0}\gets M_{r-k+1,p-s_0}\to M_{r-k+2,s_0}\gets\ldots\gets M_{r+k-1,p-s_0}\to M_{r+k,s_0}
\end{array}
\label{2r>2k}
\ee
By construction, the associated Virasoro characters satisfy
\be
 \chit[Q^{\to}_{r,s}](q)= \chit[Q^{\gets}_{r,s}](q)=\chit[Q_{r,s}](q)
\ee
Letting
\be
 \ch_{r,s}(q)=\chit[M_{r,s}](q)
\ee
denote the character of the irreducible module $M_{r,s}$, we thus have
\bea
 \chit[Q_{r,s}](q)&=&\sum_{j=0}^{\min(2r-1,2k)}\ch_{r+k-j,(-1)^j s_0+(1-(-1)^j)p/2}(q)\nn
 &=&\sum_{j=|r-k|+1,\,\mathrm{by}\,2}^{r+k-1}\ch_{j,p-s_0}(q)
  +\sum_{j=|r-k-1|+1,\,\mathrm{by}\,2}^{r+k}\ch_{j,s_0}(q)
\label{Qchar}
\eea

The range for $s_0$ in (\ref{s}) can be extended from $\mathbb{Z}_{1,p-1}$ to 
$\mathbb{Z}_{0,p-1}$ such that $s$ can be any positive integer
$s\in\mathbb{N}$ (where we exclude $s_0=k=0$ for which $s=0$).
For $s_0=0$, the structure diagrams associated with $Q^\to_{r,s}$ and $Q^\gets_{r,s}$
are separable (degenerate) and the modules are fully reducible
\be
 Q_{r,kp}^\to=Q_{r,kp}^\gets=Q_{r,kp}=\bigoplus_{j=|r-k|+1,\,\mathrm{by}\,2}^{r+k-1}M_{j,p}
\label{s00}
\ee
It is noted that this decomposition is symmetric in $r$ and $k$.
It is also noted that, for $k=0$, the finitely-generated
Feigin-Fuchs modules associated with $Q_{r,s}$ are irreducible as we have
\be
 Q_{r,s_0}^\to=Q_{r,s_0}^\gets=Q_{r,s_0}=M_{r,s_0}
\label{k0}
\ee

{}From~\cite{RP0707}, we know that the Kac representation $(r,s)$ is irreducible for $s\leq p$
and fully reducible for $s=kp$, with the set of irreducible modules denoted by
\be
 \Jc^\mathrm{Irr}=\{(r,s);\, r\in\mathbb{N},\,s\in\mathbb{Z}_{1,p}\}
\label{Jc}
\ee
A conjecture for the structure of the remaining Kac representations 
was presented in~\cite{Ras1012}. For general $(r,s)$ with $s$ given in (\ref{s}), it was thus
proposed that 
\be
 (r,s)=\begin{cases} Q^{\to}_{r,s},\ \ &2r-1<2k
  \\[.2cm]
  Q^{\gets}_{r,s},\ \ &2r-1>2k
 \end{cases}
\label{emb}
\ee
Here we adopt this assumption, but will return to it in Section \ref{SecFusVir}.
The associated Virasoro characters are denoted by $\chit_{r,s}(q)$ and by construction given by
\be
 \chit_{r,s}(q)=\chit[Q_{r,s}](q)
\ee
We note that the irreducibility of $(r,s)$ for $s\leq p$ corresponds to (\ref{k0}),
while the fully reducibility of $(r,kp)$ corresponds to (\ref{s00}).
The symmetry in $r$ and $k$ in (\ref{s00}) corresponds to the identification
$(k,rp)\equiv(r,kp)$ which reduces to the identification $(1,rp)\equiv(r,p)$
of irreducible modules. This justifies the choice of notation in (\ref{Jc}).

\subsection{Contragredient Kac representations}

In all cases (\ref{2r<2k}) and (\ref{2r>2k}), 
the finitely-generated Feigin-Fuchs modules $Q^{\to}_{r,s}$ and $Q^{\gets}_{r,s}$ 
are {\em contragredient} to each other
where the contragredient module $A^\ast$ to a module $A$ is obtained by reversing all 
structure arrows between its irreducible subfactors (subquotients). 
It follows, in particular, that $\chit[A^\ast](q)=\chit[A](q)$ and that $A^{\ast\ast}=A$.
Following~\cite{Ras1012}, the contragredient Kac representations are introduced
as 
\be
 (r,s)^\ast=\begin{cases} Q^{\gets}_{r,s},\ \ &2r-1<2k
  \\[.2cm]
  Q^{\to}_{r,s},\ \ &2r-1>2k
 \end{cases}
\label{embast}
\ee
whose Virasoro characters $\chit^\ast_{r,s}(q)=\chit[(r,s)^\ast](q)$ are given by 
$\chit^\ast_{r,s}(q)=\chit_{r,s}(q)$.
We note that $(r,s)^\ast=(r,s)$ if and only if $(r,s)$ is fully reducible, that is,
\be
 (r,s)^\ast=(r,s)\quad\Longleftrightarrow\quad s\in\mathbb{Z}_{1,p-1}\cup p\mathbb{N}
\ee

\subsection{Rank-2 and projective modules}

The infinite family 
\be
 \{\R_r^b\,;\, r\in\mathbb{N},\,b\in\mathbb{Z}_{1,p-1}\}
\label{r2}
\ee
of reducible yet indecomposable modules of rank 2 arises
from repeated fusion of {\em irreducible} Kac representations~\cite{RP0707}.
The rank-2 module $\R_r^b$ is characterized by the structure diagram
\psset{unit=.25cm}
\setlength{\unitlength}{.25cm}
\be
\mbox{}
\hspace{-1cm}
 \mbox{
 \begin{picture}(13,6)(0,3.5)
    \unitlength=1cm
  \thinlines
\put(-1.8,1){$\R_1^b:$}
\put(1,2){$M_{2,b}$}
\put(-.4,1){$M_{1,p-b}$}
\put(2,1){$M_{1,p-b}$}
\put(1.05,1){$\longleftarrow$}
\put(1.65,1.5){$\nwarrow$}
\put(0.65,1.5){$\swarrow$}
\put(3.5,1){,}
 \end{picture}
}
\hspace{3cm}
 \mbox{
 \begin{picture}(13,6)(0,3,5)
    \unitlength=1cm
  \thinlines
\put(-1.8,1){$\R_r^b:$}
\put(0.8,2){$M_{r+1,b}$}
\put(-0.4,1){$M_{r,p-b}$}
\put(2,1){$M_{r,p-b}$}
\put(0.8,0){$M_{r-1,b}$}
\put(1.05,1){$\longleftarrow$}
\put(1.65,1.5){$\nwarrow$}
\put(0.65,1.5){$\swarrow$}
\put(1.65,0.5){$\swarrow$}
\put(0.65,0.5){$\nwarrow$}
\put(3.5,1){$,\qquad r\in\mathbb{Z}_{\geq2}$}
 \end{picture}
}
\label{Remb}
\\[0.8cm]
\ee
It is noted that the rank-2 modules are all self-contragredient
\be
 (\R_r^b)^\ast=\R_r^b
\ee
The Virasoro character of the rank-2 module $\R_r^b$ follows from the structure diagram 
(\ref{Remb}) and is given by
\be
 \chit[\R_r^b](q)
  =(1-\delta_{r,1})\ch_{r-1,b}(q)+2\ch_{r,p-b}(q)+\ch_{r+1,b}(q)
\ee
According to the fusion algebra conjectured in~\cite{Ras1012} and reviewed in 
Appendix~\ref{AppFusVir}, no additional rank-2 modules nor higher-rank modules are
generated from repeated fusion of the {\em full} set of Kac representations $(r,s)$
and contragredient Kac representations $(r,s)^\ast$.

These rank-2 modules are all projective modules, but not the only projective modules in 
the model.
The Kac representations $(1,rp)\equiv(r,p)$ are both irreducible and projective as modules
over the Virasoro algebra. It is thus convenient to introduce the alternative notation
\be
 \R_r^0\equiv(1,rp)\equiv(r,p)
\label{R0}
\ee
allowing us to write the set of projective modules as
\be
 \Jc^\mathrm{Proj}=\{\R_r^b\,;\, r\in\mathbb{N},\,b\in\mathbb{Z}_{0,p-1}\}
\label{JProj}
\ee

\subsection{Fusion algebras}
\label{SecFusVir}

There are infinitely many fusion (sub)algebras associated with ${\cal LM}(1,p)$. 
Results on the corresponding fusion rules can be found 
in~\cite{GK9604,Flohr9605}.
The {\em fundamental fusion algebra}~\cite{RP0707}
\be
 \big\langle (1,1),(2,1),(1,2)\big\rangle
\label{fund}
\ee
in particular, is generated from the two fundamental Kac representations $(2,1)$ and $(1,2)$
in addition to the identity $(1,1)$.
This fusion algebra involves all the irreducible Kac representations
and all the rank-2 representations (\ref{r2}).
On the other hand, no reducible yet indecomposable Kac representations 
arise as the result of repeated fusion of the fundamental Kac representations.
The set of indecomposable modules partaking in the fundamental fusion algebra is
simply given by
\be
 \Jc^\mathrm{Fund}=\Jc^\mathrm{Irr}\cup\Jc^\mathrm{Proj}
\ee
This is not written as a disjoint union of sets since the modules (\ref{R0}) are both irreducible 
and projective.

The {\em Kac fusion algebra}
\be
 \big\langle(r,s);\, r,s\in\mathbb{N}\big\rangle
\ee
is generated by repeated fusion of the full set of Kac representations, where
\be
 (r,s)=(r,1)\otimes(1,s),\qquad \R_r^b=(r,1)\otimes\R_1^b
\label{r11s}
\ee
A concrete conjecture for this fusion algebra was presented in~\cite{Ras1012}.
It was subsequently demonstrated in~\cite{Ras1012} that this conjectured fusion algebra
is generated by repeated fusion of four Kac representations
\be
 \big\langle(r,s);\, r,s\in\mathbb{N}\big\rangle
  =\big\langle(1,1),(2,1),(1,2),(1,p+1)\big\rangle
\ee
The set of distinct, indecomposable modules partaking in this {\em Kac fusion algebra} is
\be
 \Jc^\mathrm{Kac}=\Jc^\mathrm{Fund}\cup\{(r,s);\,r\in\mathbb{N},
   s\in\mathbb{N}\setminus\big(\mathbb{Z}_{1,p-1}\cup p\mathbb{N}\big)\}
\ee
here written as a disjoint union of sets.
It was also found that the fusion algebra generated by the contragredient Kac representations
is isomorphic to the Kac fusion algebra, that is,
\be
 \big\langle(r,s)^\ast;\,r,s\in\mathbb{N}\big\rangle
 =\big\langle(1,1)^\ast,(2,1)^\ast,(1,2)^\ast,(1,p+1)^\ast\big\rangle
 \simeq\big\langle\Jc^\mathrm{Kac}\big\rangle
\label{KacKac}
\ee
where
\be
 A^\ast=A\quad \mathrm{if}\quad A\in\Jc^\mathrm{Fund},\qquad\qquad
 A^\ast\neq A\quad \mathrm{if}\quad A\in\Jc^\mathrm{Kac}\setminus\Jc^\mathrm{Fund}
\ee
 
The {\em contragrediently-extended Kac fusion algebra}
\be
 \big\langle(r,s),(r,s)^\ast;\,r,s\in\mathbb{N}\big\rangle
 =\big\langle(1,1),(2,1),(1,2),(1,p+1),(1,p+1)^\ast\big\rangle
\label{Kac5}
\ee
is generated by repeated fusion of Kac representations {\em and} contragredient Kac
representations.
As indicated, it is actually generated by repeated fusion of five modules only.
Three of these five modules are self-contragredient:
$(1,1)^\ast=(1,1)$, $(2,1)^\ast=(2,1)$ and $(1,2)^\ast=(1,2)$.
The set of distinct, indecomposable modules partaking in this fusion algebra is
\be
 \Jc^\mathrm{Cont}=\Jc^\mathrm{Kac}\cup\{(r,s)^\ast;\,r\in\mathbb{N},\,
   s\in\mathbb{N}\setminus\big(\mathbb{Z}_{1,p-1}\cup p\mathbb{N}\big)\}
\ee
here written as a disjoint union of sets.
The corresponding fusion rules are reviewed in Appendix~\ref{AppFusVir}.

The set of projective modules $\Jc^\mathrm{Proj}$
generates an {\em ideal} of the contragrediently-extended Kac fusion algebra and hence 
of the Kac fusion algebra itself as well as of the fundamental fusion algebra.
We furthermore observe that, as a factor in a fusion product, a projective module is 
{\em insensitive} to the decomposability properties of the other fusion factor. That is,
\be
 \R_r^b\otimes A=\R_r^b\otimes\big(\bigoplus_n M_n\big),\qquad 
  \R_r^b\in\Jc^\mathrm{Proj}
\label{RA}
\ee
where $\bigoplus_n M_n$
is the direct sum of the irreducible subfactors (subquotients) of the module $A$.
By construction, we thus have
\be
 \chit[A](q)=\sum_n\chit[M_n](q)
\ee
It follows from (\ref{RA}) that fusion by the projective module $\R_r^b$ is an {\em exact functor}
for all $\R_r^b\in\Jc^\mathrm{Proj}$.

As discussed in~\cite{Ras1012}, the lattice approach to the logarithmic minimal models
seems incapable of distinguishing between the family of Kac representations
and the family of contragredient Kac representations. The two families generate isomorphic
fusion algebras (\ref{KacKac}) and the corresponding characters
are identical $\chit_{r,s}(q)=\chit^\ast_{r,s}(q)$. It was simply asserted
in~\cite{Ras1012} that the Yang-Baxter integrable boundary conditions likewise denoted by $(r,s)$
in the lattice approach~\cite{PRZ0607,RP0707} are associated with the Kac representations,
even though they, a priori, could be associated with the contragredient Kac 
representations. We will return to this issue in Section~\ref{SecContFus}
when discussing fusion in the $\Wc$-extended picture.

\section{$\Wc$-extended logarithmic minimal model ${\cal WLM}(1,p)$}
\label{SecW}

\subsection{Integrable boundary conditions and $\Wc$-extended modules}

It was found in~\cite{PRR0803} that the $\Wc$-extended vacuum boundary condition can be 
constructed by fusing three $r$-type integrable seams to the boundary
\be
 \ketw{1,1}:=\lim_{n\to\infty}(2n-1,1)\otimes(2n-1,1)\otimes(2n-1,1)
  =\bigoplus_{n=1}^\infty \,(2n-1)\,(2n-1,1)
\label{11W}
\ee
thereby ensuring that the $\Wc$-extended 
vacuum boundary condition is a solution to the boundary Yang-Baxter equation.
The corresponding $\Wc$-extended module $\ketw{1,1}$ is indecomposable
(in fact, irreducible) with respect to the $W$-algebra $\Wc(p)$, but decomposable
with respect to the Virasoro algebra. Its decomposition in terms of indecomposable
Virasoro modules appears as the last expression in (\ref{11W}).
These Virasoro modules are all irreducible.

Using the stability properties~\cite{PRR0803}
\bea
 (2m-1,s)\otimes \ketw{1,1}&=&(2m-1)\,\Big(\bigoplus_{n=1}^\infty \,(2n-1)\,(2n-1,s)\Big)\nn
 (2m,s)\otimes \ketw{1,1}&=&2m\,\Big(\bigoplus_{n=1}^\infty \,2n\,(2n,s)\Big)\nn
 \Rc_{2m-1}^b\otimes\ketw{1,1}&=&(2m-1)\,\Big(\bigoplus_{n=1}^\infty \,(2n-1)\,\Rc_{2n-1}^b\Big)
     \nn
 \Rc_{2m}^b\otimes\ketw{1,1}&=&2m\,\Big(\bigoplus_{n=1}^\infty \,2n\,\Rc_{2n}^b\Big)
\label{prop11gen}
\eea
for $s\in\mathbb{Z}_{1,p}$, $b\in\mathbb{Z}_{1,p-1}$ and $m\in\mathbb{N}$, one
can identify integrable boundary conditions corresponding to the $2p$ $\Wc$-irreducible modules
\bea
 \ketw{1,s}&\!\!:=\!\!&(1,s)\otimes \ketw{1,1}\;=\;\bigoplus_{n=1}^\infty \,(2n-1)\,(2n-1,s)\nn
 \ketw{2,s}&\!\!:=\!\!&\half (2,s)\otimes \ketw{1,1}\;=\;\bigoplus_{n=1}^\infty \,2n\,(2n,s)
\label{ketw}
\eea
and the $2p-2$ $\Wc$-reducible yet $\Wc$-indecomposable rank-2 modules
\bea
 \Rh_{1}^b
  &\!\!:=\!\!&\Rc_{1}^b\otimes\ketw{1,1}\;=\;\bigoplus_{n=1}^\infty \,(2n-1)\,\Rc_{2n-1}^b\nn
 \Rh_{2}^b
  &\!\!:=\!\!&\half\Rc_{2}^b\otimes\ketw{1,1}\;=\;\bigoplus_{n=1}^\infty \,2n\,\Rc_{2n}^b
\label{ketwR}
\eea
The set $\Jc^\mathrm{Irr}_\Wc$ of $\Wc$-irreducible modules is given by
\be
 \Jc_\Wc^\mathrm{Irr}=\{\ketw{r,s};\,r\in\mathbb{Z}_{1,2},\,s\in\mathbb{Z}_{1,p}\}
 =\{\Mh_{r,s};\,r\in\mathbb{Z}_{1,2},\,s\in\mathbb{Z}_{1,p}\}
\ee
where we have introduced the notation $\Mh_{r,s}$ to denote a $\Wc$-irreducible
module:
\be
 \Mh_{r,s}=\ketw{r,s}\quad \mathrm{if}\quad r\in\mathbb{Z}_{1,2},\,s\in\mathbb{Z}_{1,p}
\label{Mh}
\ee
The structure diagrams of the rank-2 modules are of the form
\psset{unit=.25cm}
\setlength{\unitlength}{.25cm}
\begin{equation}
 \mbox{
 \begin{picture}(20,8.5)(-6,1.1)
    \unitlength=1cm
  \thinlines
\put(-2.3,1){$\Rh_r^b:$}
\put(0.95,2){$\Mh_{2\cdot r,b}$}
\put(-0.35,1){$\Mh_{r,p-b}$}
\put(2,1){$\Mh_{r,p-b}$}
\put(1,0){$\Mh_{2\cdot r,b}$}
\put(1.05,1){$\longleftarrow$}
\put(1.65,1.5){$\nwarrow$}
\put(0.65,1.5){$\swarrow$}
\put(1.65,0.5){$\swarrow$}
\put(0.65,0.5){$\nwarrow$}
 \end{picture}
}
\label{Rdiagram}
\end{equation} 
Introducing 
\be
 \Rh_{r}^0\equiv\ketw{r,p}=\Mh_{r,p},\qquad r\in\mathbb{Z}_{1,2}
\label{Rh0}
\ee
the set of $\Wc$-projective modules is
\be
 \Jc_\Wc^\mathrm{Proj}=\{\Rh_r^b;\,r\in\mathbb{Z}_{1,2},\,b\in\mathbb{Z}_{0,p-1}\}
\ee
It is noted that the two modules in (\ref{Rh0}) are both $\Wc$-irreducible and $\Wc$-projective,
and that they are the only such modules.
The Virasoro characters of the $\Wc$-indecomposable modules (\ref{ketw}) and (\ref{ketwR}) 
follow readily from the indicated decompositions in terms of Virasoro modules.

The work~\cite{PRR0803}, in which the lattice construction of the
$\Wc$-extended modules (\ref{ketw}) and (\ref{ketwR})
first appeared, was focused on the construction of $\Wc$-irreducible and $\Wc$-projective
modules and on their fusion properties. The ensuing fusion algebra thus corresponds
to a lift of the fundamental fusion algebra $\langle\Jc^\mathrm{Fund}\rangle$ to
the {\em $\Wc$-extended fundamental fusion algebra}
\be
 \big\langle\Jc_\Wc^\mathrm{Fund}\big\rangle
  =\big\langle\ketw{1,1},\ketw{2,1},\ketw{1,2}\big\rangle
  =\big\langle\Mh_{1,1},\Mh_{2,1},\Mh_{1,2}\big\rangle
\label{WFund}
\ee
The set of $\Wc$-modules partaking in this fusion algebra is
\be
 \Jc_\Wc^\mathrm{Fund}=\Jc_\Wc^\mathrm{Irr}\cup\Jc_\Wc^\mathrm{Proj}
\ee
This is not a disjoint union of sets since the modules in (\ref{Rh0}) are both $\Wc$-irreducible 
and $\Wc$-projective.

For later convenience, we introduce the redundant notation
\be
 \ketw{0,s}\equiv\ketw{r,0}\equiv\Rh_0^b\equiv0
\ee

\subsection{$\Wc$-extended Kac representations}

With the recent advances~\cite{Ras1012} in the understanding of the general
structure and fusion of Kac representations reviewed in Section \ref{SecLM1p},
we now turn to the construction of the corresponding lift to the $\Wc$-extended picture.

Since
\be
 (r,s)=(r,1)\otimes(1,s)
\label{rsr11s}
\ee
it is readily seen that the validity of the stability properties (\ref{prop11gen}) extends from 
$s\in\mathbb{Z}_{1,p}$ to $s\in\mathbb{N}$.
For every pair of positive Kac labels $r,s\in\mathbb{N}$,
we can therefore define the $\Wc$-extended Kac representation $\ketw{r,s}$ by
\be
 \ketw{r,s}:=\tfrac{1}{r}(r,s)\otimes\ketw{1,1}
\label{ketwrs}
\ee
Occasionally, we will refer to these $\Wc$-extended Kac representations
simply as $\Wc$-Kac representations.
It follows from (\ref{prop11gen}) that
\be
 \ketw{r,s}=\ketw{1\cdot r,s}
\label{r1r}
\ee
and since $1\cdot r\in\mathbb{Z}_{1,2}$,
it thus suffices to define $\ketw{r,s}$ for $r\in\mathbb{Z}_{1,2}$.
Likewise, it is also sufficient to define the ${\cal W}$-extended rank-2 modules
$\Rh_r^b$ for $r\in\mathbb{Z}_{1,2}$ only, since
\be
 \Rh_r^b:=\tfrac{1}{r}\R_r^b\otimes\ketw{1,1}=\Rh_{1\cdot r}^b,\qquad r\in\mathbb{N},
  \quad b\in\mathbb{Z}_{1,p-1}
\ee
In the following, we therefore let $r\in\mathbb{Z}_{1,2}$.

Recalling $(1,kp)\equiv(k,p)$, we immediately obtain the 
identifications  
\be
 \ketw{r,kp}\equiv k\ketw{r\cdot k,p}=k\Mh_{r\cdot k,p}
\ee
of ${\cal W}$-extended modules. It follows that the modules $\ketw{r,kp}$ are fully reducible.

Based on the conjecture (\ref{emb}) for the structure of the Kac representations, we
find that the $\Wc$-Kac representation $\ketw{r,s}$ is the
finitely-generated $\Wc$-extended Feigin-Fuchs module
\be
 \ketw{r,s_0+kp}=\Qh^\gets_{r,s_0+kp},\qquad 
    s_0\in\mathbb{Z}_{0,p-1},\quad k\in\mathbb{N}_0
\label{rs0}
\ee
whose structure diagram is given by
\be
 \Qh^\gets_{r,s_0+kp}:\qquad
  \underbrace{\Mh_{2\cdot r\cdot k,s_0}\gets\Mh_{r\cdot k,p-s_0}\to\Mh_{2\cdot r\cdot k,s_0}
    \gets\ldots\gets 
  \Mh_{r\cdot k,p-s_0}\to\Mh_{2\cdot r\cdot k,s_0}}_{\#\,=\,2k+1}
\label{Qrs0}
\ee
Before justifying this claim, we note that for $k=0$, it correctly reduces to
\be
 \ketw{r,s_0}=\Qh^\gets_{r,s_0}=\Mh_{r,s_0}
\ee
while for $s_0=0$, it correctly reduces to $\ketw{r,kp}=k\Mh_{r\cdot k,p}$.
The $\Wc$-extended Kac characters following from (\ref{rs0}) and (\ref{Qrs0})
are given by
\bea
 \chih_{r,s_0+kp}(q)&=&\chit[\ketw{r,s_0+kp}](q)
    =k\,\chih_{r\cdot k,p-s_0}(q)+(k+1)\chih_{2\cdot r\cdot k,s_0}(q)
   \nonumber\\[.3cm]
 &=&\begin{cases}
  \displaystyle{\,k\sum_{n\in\mathbb{N}}(2n-1)\ch_{2n-1,p-s_0}(q)
   +(k+1)\sum_{n\in\mathbb{N}}2n\,\ch_{2n,s_0}(q)},\qquad r\cdot k=1
  \\[.7cm]
  \displaystyle{\,k\sum_{n\in\mathbb{N}}2n\,\ch_{2n,p-s_0}(q)
   +(k+1)\sum_{n\in\mathbb{N}}(2n-1)\ch_{2n-1,s_0}(q)},\qquad r\cdot k=2
  \end{cases}
\label{chihs0}
\eea
Our argument for the structure (\ref{rs0}) of the $\Wc$-Kac representation
$\ketw{r,s}$ is based on 
\begin{itemize}
\item[$(i)$] the conjectured structure diagrams (\ref{emb}) of the Kac representations
appearing in the decomposition of $\ketw{r,s}$ in terms of indecomposable Virasoro
modules; 
\item[$(ii)$] the assertion that the $\Wc$-indecomposable module $\ketw{r,s}$
can be described by a structure diagram linking $\Wc$-irreducible modules only.
\end{itemize}
To determine the structure diagram of $\ketw{r,s}$, we thus have to 
`add' or `glue together' the infinite sequence of structure diagrams associated with 
the participating Kac representations to form a single structure diagram 
involving $\Wc$-irreducible modules only.

First, for 
\be
 s=s_0+kp,\qquad s_0\in\mathbb{Z}_{0,p-1},\quad k\in\mathbb{N}_0
\label{s0}
\ee
we see that the $\Wc$-Kac representation $\ketw{r,s}$ as defined in (\ref{ketwrs}) decomposes as
\bea
 \ketw{1,s}&=&\Big(\bigoplus_{n=1}^{\lceil\frac{k}{2}\rceil}(2n-1)Q_{2n-1,s}^\to\Big)
   \oplus\Big(\bigoplus_{n=\lfloor\frac{k+3}{2}\rfloor}^{\infty}(2n-1)Q_{2n-1,s}^\gets\Big)\nn
 \ketw{2,s}&=&\Big(\bigoplus_{n=1}^{\lceil\frac{k-1}{2}\rceil}2n\,Q_{2n,s}^\to\Big)
   \oplus\Big(\bigoplus_{n=\lfloor\frac{k+2}{2}\rfloor}^{\infty}2n\,Q_{2n,s}^\gets\Big)
\label{1s2s}
\eea
in terms of indecomposable Virasoro modules. Here we have used the
conjectured structure of the Kac representations discussed in Section
\ref{SecKacRep}.
Using (\ref{Qchar}) and the sum formula
\be
 \sum_{j=|n-n'|+1,\,\mathrm{by}\,2}^{n+n'-1}j=nn'
\label{sumj}
\ee
it is verified that the Virasoro characters of (\ref{1s2s})
agree with the Virasoro characters (\ref{chihs0}) of the proposed
Feigin-Fuchs structures (\ref{rs0}).
The appearance of the indecomposable Virasoro modules
$Q_{2n-1,s}^\gets$ or $Q_{2n,s}^\gets$ in (\ref{1s2s}) requires that similar indecomposable
structures are present in the ambient $\Wc$-Kac representation as well.
Following assertion $(ii)$ above, we are thus led to the conjecture (\ref{rs0}).

For ${\cal WLM}(1,p)$, $p\in\mathbb{Z}_{2,5}$, Kac tables of the conformal weights $\D_{r,s}$
of the $\Wc$-irreducible modules $\Mh_{r,s}$ (\ref{Mh}) over the triplet $W$-algebra $\Wc(p)$
appear in Figure~\ref{FigKacTable}.
\psset{unit=.9cm}
\begin{figure}
\begin{center}
\begin{pspicture}(0,0)(2,2)
\rput(1,2.5){${\cal WLM}(1,2)$}
\psframe[linewidth=0pt,fillstyle=solid,fillcolor=lightlightblue](0,0)(2,2)
\multiput(0,0)(0,2){1}{\psframe[linewidth=0pt,fillstyle=solid,fillcolor=midblue](0,1)(2,2)}
\psgrid[gridlabels=0pt,subgriddiv=1]
\rput(.5,1.5){$-\frac 18$}\rput(1.5,1.5){$\frac 38$}
\rput(.5,.5){$0$}\rput(1.5,.5){$1$}
{\color{blue}
\rput(.5,-.5){$1$}
\rput(1.5,-.5){$2$}
\rput(-.5,.5){$1$}
\rput(-.5,1.5){$2$}
}
\end{pspicture}\qquad\qquad\qquad
\begin{pspicture}(0,0)(2,3)
\rput(1,3.5){${\cal WLM}(1,3)$}
\psframe[linewidth=0pt,fillstyle=solid,fillcolor=lightlightblue](0,0)(2,3)
\multiput(0,1)(0,3){1}{\psframe[linewidth=0pt,fillstyle=solid,fillcolor=midblue](0,1)(2,2)}
\psgrid[gridlabels=0pt,subgriddiv=1]
\rput(.5,2.5){$-\frac 13$}\rput(1.5,2.5){$\frac 5{12}$}
\rput(.5,1.5){$-\frac 14$}\rput(1.5,1.5){$1$}
\rput(.5,.5){$0$}\rput(1.5,.5){$\frac 74$}
{\color{blue}
\rput(.5,-.5){$1$}
\rput(1.5,-.5){$2$}
\rput(-.5,.5){$1$}
\rput(-.5,1.5){$2$}
\rput(-.5,2.5){$3$}
}
\end{pspicture}\qquad\qquad\qquad
\begin{pspicture}(0,0)(2,4)
\rput(1,4.5){${\cal WLM}(1,4)$}
\psframe[linewidth=0pt,fillstyle=solid,fillcolor=lightlightblue](0,0)(2,4)
\multiput(0,2)(0,4){1}{\psframe[linewidth=0pt,fillstyle=solid,fillcolor=midblue](0,1)(2,2)}
\psgrid[gridlabels=0pt,subgriddiv=1]
\rput(.5,3.5){$-\frac{9}{16}$}\rput(1.5,3.5){$\frac{7}{16}$}
\rput(.5,2.5){$-\frac{1}{2}$}\rput(1.5,2.5){$1$}
\rput(.5,1.5){$-\frac{5}{16}$}\rput(1.5,1.5){$\frac{27}{16}$}
\rput(.5,.5){$0$}\rput(1.5,.5){$\frac{5}{2}$}
{\color{blue}
\rput(.5,-.5){$1$}
\rput(1.5,-.5){$2$}
\rput(-.5,.5){$1$}
\rput(-.5,1.5){$2$}
\rput(-.5,2.5){$3$}
\rput(-.5,3.5){$4$}
}
\end{pspicture}\qquad\qquad\qquad
\begin{pspicture}(0,0)(2,5)
\rput(1,5.5){${\cal WLM}(1,5)$}
\psframe[linewidth=0pt,fillstyle=solid,fillcolor=lightlightblue](0,0)(2,5)
\multiput(0,3)(0,5){1}{\psframe[linewidth=0pt,fillstyle=solid,fillcolor=midblue](0,1)(2,2)}
\psgrid[gridlabels=0pt,subgriddiv=1]
\rput(.5,4.5){$-\frac{4}{5}$}\rput(1.5,4.5){$\frac{9}{20}$}
\rput(.5,3.5){$-\frac{3}{4}$}\rput(1.5,3.5){$1$}
\rput(.5,2.5){$-\frac{3}{5}$}\rput(1.5,2.5){$\frac{33}{20}$}
\rput(.5,1.5){$-\frac{7}{20}$}\rput(1.5,1.5){$\frac{12}{5}$}
\rput(.5,.5){$0$}\rput(1.5,.5){$\frac{13}{4}$}
{\color{blue}
\rput(.5,-.5){$1$}
\rput(1.5,-.5){$2$}
\rput(-.5,.5){$1$}
\rput(-.5,1.5){$2$}
\rput(-.5,2.5){$3$}
\rput(-.5,3.5){$4$}
\rput(-.5,4.5){$5$}
}
\end{pspicture}
\end{center}
\caption{Kac tables of conformal weights 
$\Delta_{r,s}$ of the $\Wc$-irreducible $\Wc$-Kac representations in ${\cal WLM}(1,p)$ where
$p\in\mathbb{Z}_{2,5}$. 
The corresponding central charges are $c=-2$, $c= -7$, $c=-25/2$ and $c=-91/5$, 
respectively. 
In a given table, the column index is $r\in\mathbb{Z}_{1,2}$, while the row index is 
$s\in\mathbb{Z}_{1,p}$.
The pair of $\Wc$-Kac representations which are both 
$\Wc$-irreducible and $\Wc$-projective are indicated by darker shadings.
}
\label{FigKacTable}
\end{figure}

\subsection{$\Wc$-extended Kac representations in ${\cal WLM}(1,2)$}

Here we illustrate the structure diagrams of the $\Wc$-indecomposable modules
given in (\ref{Qrs0}) for the logarithmic minimal model ${\cal WLM}(1,2)$. 
In the following, we let $k$ denote a non-negative integer.
The structure diagrams are 
\psset{unit=.5cm}
\setlength{\unitlength}{.5cm}
\bea
 \ketw{1,4k-1}:\qquad
\begin{pspicture}(0,.5)(15,2)
 \rput(0,1.5){$1$}
 \rput(0.75,0.7){$\nwarrow$}
 \rput(1.5,0){$0$}
 \rput(2.2,0.7){$\nearrow$}
 \rput(2.95,1.5){$1$}
 \rput(3.65,0.7){$\nwarrow$}
 \rput(4.37,0){$0$}
 \rput(5.1,0.7){$\nearrow$}
 \rput(6.4,1){$\ldots$}
 \rput(7.65,0.7){$\nwarrow$}
 \rput(8.4,0){$0$}
 \rput(9.1,0.7){$\nearrow$}
 \rput(9.8,1.5){$1$}
 \rput(13.35,1.5){\#\,=\,$2k$}
 \rput(14,0){\#\,=\,$2k-1$}
\end{pspicture} 
\nonumber
\\[.15cm]
\\[.15cm]
 \ketw{2,4k-1}:\qquad
\begin{pspicture}(0,.5)(15,2)
 \rput(0,0){$0$}
 \rput(0.75,0.7){$\swarrow$}
 \rput(1.5,1.5){$1$}
 \rput(2.2,0.7){$\searrow$}
 \rput(2.95,0){$0$}
 \rput(3.65,0.7){$\swarrow$}
 \rput(4.37,1.5){$1$}
 \rput(5.1,0.7){$\searrow$}
 \rput(6.4,0){$\ldots$}
 \rput(7.65,0.7){$\swarrow$}
 \rput(8.4,1.5){$1$}
 \rput(9.1,0.7){$\searrow$}
 \rput(9.8,0){$0$}
 \rput(14,1.5){\#\,=\,$2k-1$}
 \rput(13.35,0){\#\,=\,$2k$}
\end{pspicture} 
\nn
\nonumber
\eea
and
\bea
 \ketw{1,4k+1}:\qquad
\begin{pspicture}(0,.5)(15,2)
 \rput(0,0){$0$}
 \rput(0.75,0.7){$\swarrow$}
 \rput(1.5,1.5){$1$}
 \rput(2.2,0.7){$\searrow$}
 \rput(2.95,0){$0$}
 \rput(3.65,0.7){$\swarrow$}
 \rput(4.37,1.5){$1$}
 \rput(5.1,0.7){$\searrow$}
 \rput(6.4,0){$\ldots$}
 \rput(7.65,0.7){$\swarrow$}
 \rput(8.4,1.5){$1$}
 \rput(9.1,0.7){$\searrow$}
 \rput(9.8,0){$0$}
 \rput(13.35,1.5){\#\,=\,$2k$}
 \rput(14,0){\#\,=\,$2k+1$}
\end{pspicture} 
\nonumber
\\[.15cm]
\\[.15cm]
 \ketw{2,4k+1}:\qquad
\begin{pspicture}(0,.5)(15,2)
 \rput(0,1.5){$1$}
 \rput(0.75,0.7){$\nwarrow$}
 \rput(1.5,0){$0$}
 \rput(2.2,0.7){$\nearrow$}
 \rput(2.95,1.5){$1$}
 \rput(3.65,0.7){$\nwarrow$}
 \rput(4.37,0){$0$}
 \rput(5.1,0.7){$\nearrow$}
 \rput(6.4,1){$\ldots$}
 \rput(7.65,0.7){$\nwarrow$}
 \rput(8.4,0){$0$}
 \rput(9.1,0.7){$\nearrow$}
 \rput(9.8,1.5){$1$}
 \rput(14,1.5){\#\,=\,$2k+1$}
 \rput(13.35,0){\#\,=\,$2k$}
\end{pspicture} 
\nn
\nonumber
\eea
where the $\Wc$-irreducible module $\Mh_{r,s}$ (\ref{Mh}) is represented by its
conformal weight $\D_{r,s}$.
For $k=1$, we thus have
\be
 \ketw{1,3}:\qquad
\begin{pspicture}(0,.5)(3,2)
 \rput(0,1.5){$1$}
 \rput(0.75,0.7){$\nwarrow$}
 \rput(1.5,0){$0$}
 \rput(2.2,0.7){$\nearrow$}
 \rput(2.95,1.5){$1$}
\end{pspicture} 
\hspace{3.15cm}
 \ketw{2,3}:\qquad
\begin{pspicture}(0,.5)(3,2)
 \rput(0,0){$0$}
 \rput(0.75,0.7){$\swarrow$}
 \rput(1.5,1.5){$1$}
 \rput(2.2,0.7){$\searrow$}
 \rput(2.95,0){$0$}
\end{pspicture} 
\label{1323}
\\[.3cm]
\ee
and
\be
 \ketw{1,5}:\qquad
\begin{pspicture}(0,.5)(5,2)
 \rput(0,0){$0$}
 \rput(0.75,0.7){$\swarrow$}
 \rput(1.5,1.5){$1$}
 \rput(2.2,0.7){$\searrow$}
 \rput(2.95,0){$0$}
 \rput(3.65,0.7){$\swarrow$}
 \rput(4.37,1.5){$1$}
 \rput(5.1,0.7){$\searrow$}
 \rput(5.85,0){$0$}
\end{pspicture} 
\hspace{2.2cm}
 \ketw{2,5}:\qquad
\begin{pspicture}(0,.5)(5,2)
 \rput(0,1.5){$1$}
 \rput(0.75,0.7){$\nwarrow$}
 \rput(1.5,0){$0$}
 \rput(2.2,0.7){$\nearrow$}
 \rput(2.95,1.5){$1$}
 \rput(3.65,0.7){$\nwarrow$}
 \rput(4.37,0){$0$}
 \rput(5.1,0.7){$\nearrow$}
 \rput(5.85,1.5){$1$}
\end{pspicture} 
\label{1525}
\\[.3cm]
\ee

\subsection{$\Wc$-extended Kac fusion algebra}

The fusion rules in the $\Wc$-extended picture are inferred from the fusion rules
in the Virasoro picture.
Letting $\hat\otimes$ denote the fusion multiplication in the $\Wc$-extended picture,
it is interpreted~\cite{PRR0803} as a limit of a rescaled fusion  
\begin{equation}
 \ketw{1,1}\, \hat\otimes\, \ketw{1,1}:=\lim_{n\to\infty}\Big(\frac{1}{(2n-1)^3}(2n-1,1)\otimes(2n-1,1)
   \otimes(2n-1,1)\otimes\ketw{1,1}\Big)=\ketw{1,1}
\end{equation}
in the Virasoro picture of the logarithmic minimal model ${\cal LM}(1,p)$.
This ensures that fusion in the extended picture has a natural implementation on the lattice.

Now, a representation $\hat{A}$ in the $\Wc$-extended picture is constructed as the integrable
boundary condition $A\otimes\ketw{1,1}$ where $A$ is some Virasoro representation in the 
logarithmic minimal model. Fusion in the extended picture is then computed as
\bea
 \hat{A}\fus\hat{B}&=&
  \big(A\otimes \ketw{1,1}\big)\fus\big(B\otimes\ketw{1,1}\big)
   =\big(A\otimes B\big)\otimes\big(\ketw{1,1}\fus\ketw{1,1}\big)\nn
  &=&\big(\bigoplus_jC_j\big)\otimes\ketw{1,1}
   \ =\ \bigoplus_j\hat{C_j}
\label{ABC}
\eea
where $A\otimes B=\bigoplus_jC_j$ is the fusion of the representations $A$ and $B$ in the
Virasoro picture.
This $\Wc$-extended fusion prescription is readily seen to be both associative and commutative.
It is also immediately verified that $\ketw{1,1}$ is the identity of the ensuing fusion algebra
\begin{equation}
 \ketw{1,1}\fus\hat{A}
  =\big((1,1)\otimes \ketw{1,1}\big)\fus\big(A\otimes\ketw{1,1}\big)
  =\big((1,1)\otimes A\big)\otimes\ketw{1,1}=\hat{A}
\end{equation}

With this $\Wc$-extended fusion prescription, it follows that the $\Wc$-Kac 
representation $\ketw{r,s}$ `separates' in much the same way (\ref{rsr11s}) as the original 
Kac representations, that is,
\bea
 \ketw{r,s}&=&\frac{1}{r}(r,s)\otimes\ketw{1,1}=[\frac{1}{r}(r,1)\otimes(1,s)]
  \otimes[\ketw{1,1}\fus\ketw{1,1}]\nn
 &=&[\frac{1}{r}(r,1)\otimes\ketw{1,1}]\fus[(1,s)\otimes\ketw{1,1}]\nn
 &=&\ketw{r,1}\fus\ketw{1,s}
\label{sep}
\eea
Using (\ref{sumj}), we also find that
\be
 \ketw{r,1}\fus\ketw{r',1}=\ketw{r\cdot r',1},\qquad
 \ketw{r,1}\fus\Rh_{r'}^b=\Rh_{r\cdot r'}^b
\ee
and hence
\be
 \ketw{r,s}\fus\ketw{r',s'}=\ketw{r\cdot r',1}\fus\big[\ketw{1,s}\fus\ketw{1,s'}\big]
\label{rsrs}
\ee
and
\be
 \Rh_{r}^{b}\fus\ketw{r',s'}=\ketw{r\cdot r',1}\fus\big[\Rh_1^{b}\fus\ketw{1,s'}\big],
  \qquad
 \Rh_r^b\fus\Rh_{r'}^{b'}=\ketw{r\cdot r',1}\fus\big[\Rh_1^b\fus\Rh_1^{b'}\big]
\label{rsR}
\ee

Based on the fusion rules~\cite{Ras1012} in the Virasoro picture,
summarized in Appendix \ref{AppFusVir}, we work out the {\em $\Wc$-extended Kac fusion algebra} 
\be
 \big\langle\ketw{r,s};\,r\in\mathbb{Z}_{1,2},\,s\in\mathbb{N}\big\rangle
\ee
generated by repeated fusion of the $\Wc$-Kac representations.
Written as a disjoint union of sets, 
the set of distinct, $\Wc$-indecomposable modules partaking in this fusion algebra is
\be
 \Jc_\Wc^\mathrm{Kac}=\{\ketw{r,s};\,r\in\mathbb{Z}_{1,2},\,s\in\mathbb{N}\setminus p\mathbb{N}\}
  \cup\Jc_{\Wc}^\mathrm{Proj}
\ee
For $r,r'\in\mathbb{Z}_{1,2}$, $b,b'\in\mathbb{Z}_{0,p-1}$ and $k,k'\in\mathbb{N}_0$, 
we find the underlying fusion rules to be given by
\bea
 \ketw{r,b+kp}\fus\ketw{r',b'+k'p}
 &=&kk'\Big(\bigoplus_\beta^{p-|b-b'|-1}\Rh_{r\cdot r'\cdot k\cdot k'}^\beta\Big)
  \oplus
  (k+k'+1)\Big(\bigoplus_\beta^{b+b'-p-1}\Rh_{r\cdot r'\cdot k\cdot k'}^\beta\Big)\nn
 &\oplus&
  (k+1)k'\Big(\bigoplus_\beta^{b-b'-1}\Rh_{2\cdot r\cdot r'\cdot k\cdot k'}^\beta\Big)
  \oplus 
  k(k'+1)\Big(\bigoplus_\beta^{b'-b-1}\Rh_{2\cdot r\cdot r'\cdot k\cdot k'}^\beta\Big)\nn
 &\oplus&
  \bigoplus_{\beta=|b-b'|+1,\ \!\mathrm{by}\ \!2}^{p-|p-b-b'|-1}\ketw{r\cdot r',\beta+(k+k')p}
\label{rbrb}
\eea
and
\bea
 \Rh_r^b\fus\ketw{r',b'+k'p}
  &=&\Big\{k'\Big(\bigoplus_\beta^{p-|b-b'|-1}\Rh_{r\cdot r'\cdot k'}^\beta\Big)
  \oplus k'\Big(\bigoplus_\beta^{|p-b-b'|-1}\Rh_{r\cdot r'\cdot k'}^\beta\Big)
  \oplus2\Big(\bigoplus_\beta^{b+b'-p-1}\Rh_{r\cdot r'\cdot k'}^\beta\Big)\nn
 &&\quad\oplus\,(k'+1)\Big(\bigoplus_{\beta=|b-b'|+1,\,\mathrm{by}\,2}^{p-|p-b-b'|-1}
   \Rh_{2\cdot r\cdot r'\cdot k'}^\beta\Big)
 \oplus2k'\Big(\bigoplus_\beta^{|b-b'|-1}\Rh_{2\cdot r\cdot r'\cdot k'}^\beta\Big)\nn
 &&\quad\oplus\,2\Big(\bigoplus_\beta^{b'-b-1}
   \Rh_{2\cdot r\cdot r'\cdot k'}^\beta\Big)\Big\}/(1+\delta_{b,0})
\label{fusRKac}
\eea
as well as the known fusion rules 
\bea
 \Rh_{r}^{b}\fus\Rh_{r'}^{b'}
  &=&2\Big(\bigoplus_{\beta}^{p-|b-b'|-1}
     \Rh_{r\cdot r'}^{\beta}
     \oplus\!\!\bigoplus_{\beta}^{|p-b-b'|-1}
     \Rh_{r\cdot r'}^{\beta}\nn
   &&\qquad\quad\oplus\!\bigoplus_{\beta}^{p-|p-b-b'|-1}
     \Rh_{2\cdot r\cdot r'}^{\beta}
     \oplus\!\bigoplus_{\beta}^{|b-b'|-1}
     \Rh_{2\cdot r\cdot r'}^{\beta}\Big)/\{(1+\delta_{b,0})(1+\delta_{b',0})\}
\label{fus}
\eea
for the subalgebra generated by the projective modules $\Rh_r^b$.
This subalgebra is actually an {\em ideal}, in accordance with the modules $\Rh_r^b$ being 
projective. The divisions in (\ref{fusRKac}) and (\ref{fus}) by $(1+\delta_{b,0})$ and 
$(1+\delta_{b',0})$ ensure that the fusion rules for $\Rh_r^0$ match those for $\ketw{r,p}$.

We observe that, as a factor in a fusion product, the $\Wc$-projective module $\Rh_r^b$
is {\em insensitive} to the indecomposable structure of the other $\Wc$-extended fusion factor, 
that is,
\bea
 \Rh_r^b\fus\Rh_{r'}^{b'}
   &=&\Rh_r^b\fus\big[2\ketw{r',p-b'}\oplus2\ketw{2\cdot r',b'}\big]\nn
 \Rh_r^b\fus\ketw{r',b'+k'p}
   &=&\Rh_r^b\fus\big[k'\ketw{r'\cdot k',p-b'}\oplus(k'+1)\ketw{2\cdot r'\cdot k',b'}\big]
\label{Winsensitive}
\eea
Here we have $b\in\mathbb{Z}_{0,p-1}$, while we set $b'\in\mathbb{Z}_{1,p-1}$ for $\Rh_{r'}^{b'}$ 
and $\ketw{r',b'+k'p}$ to be reducible yet indecomposable.
Similar to the situation in the Virasoro case, the insensitivity properties (\ref{Winsensitive}) 
imply that fusion by $\Rh_r^b$ in $\langle\Jc^\mathrm{Kac}_\Wc\rangle$
is an {\em exact functor} for all $\Rh_r^b\in\Jc^\mathrm{Proj}_\Wc$.

\subsection{Contragredient modules and their fusion properties}
\label{SecContFus}

As in the Virasoro picture, we introduce the contragredient module to each $\Wc$-Kac
representation $\ketw{r,s}$ by reversing the arrows in the corresponding structure diagram
(\ref{rs0}) and (\ref{Qrs0}). 
For $r\in\mathbb{N}$, $s_0\in\mathbb{Z}_{0,p-1}$ and $k\in\mathbb{N}_{0}$,
we thus have
\be
 \ketwa{r,s_0+kp}=\Qh^\to_{r,s_0+kp}
\ee
whose structure diagram is given by
\be
 \Qh^\to_{r,s_0+kp}:\qquad
  \underbrace{\Mh_{2\cdot r\cdot k,s_0}\to\Mh_{r\cdot k,p-s_0}\gets\Mh_{2\cdot r\cdot k,s_0}
     \to\ldots\to
  \Mh_{r\cdot k,p-s_0}\gets\Mh_{2\cdot r\cdot k,s_0}}_{\#\,=\,2k+1}
\ee
The corresponding character is denoted by $\chih^\ast_{r,s}(q)=\chit[\ketwa{r,s}](q)$, and
it follows that $\ketwa{r,s}=\ketw{r,s}$ if and only if $\ketw{r,s}$ is fully reducible, that is,
\be
 \ketwa{r,s}=\ketw{r,s}\quad\Longleftrightarrow\quad s\in\mathbb{Z}_{1,p-1}\cup p\mathbb{N}
\label{KastK}
\ee
As in the case of the $\Wc$-Kac representations themselves, we may restrict
our considerations of $\ketwa{r,s}$ to $r\in\mathbb{Z}_{1,2}$ since $\ketwa{r,s}=\ketwa{r\cdot1,s}$.
For ${\cal WLM}(1,2)$, the contragredient modules to the ones described explicitly in
(\ref{1323}) and (\ref{1525}) are
\be
 \ketwa{1,3}:\qquad
\begin{pspicture}(0,.5)(3,2)
 \rput(0,1.5){$1$}
 \rput(0.75,0.7){$\searrow$}
 \rput(1.5,0){$0$}
 \rput(2.2,0.7){$\swarrow$}
 \rput(2.95,1.5){$1$}
\end{pspicture} 
\hspace{3.15cm}
 \ketwa{2,3}:\qquad
\begin{pspicture}(0,.5)(3,2)
 \rput(0,0){$0$}
 \rput(0.75,0.7){$\nearrow$}
 \rput(1.5,1.5){$1$}
 \rput(2.2,0.7){$\nwarrow$}
 \rput(2.95,0){$0$}
\end{pspicture} 
\\[.3cm]
\ee
and
\be
 \ketwa{1,5}:\qquad
\begin{pspicture}(0,.5)(5,2)
 \rput(0,0){$0$}
 \rput(0.75,0.7){$\nearrow$}
 \rput(1.5,1.5){$1$}
 \rput(2.2,0.7){$\nwarrow$}
 \rput(2.95,0){$0$}
 \rput(3.65,0.7){$\nearrow$}
 \rput(4.37,1.5){$1$}
 \rput(5.1,0.7){$\nwarrow$}
 \rput(5.85,0){$0$}
\end{pspicture} 
\hspace{2.2cm}
 \ketwa{2,5}:\qquad
\begin{pspicture}(0,.5)(5,2)
 \rput(0,1.5){$1$}
 \rput(0.75,0.7){$\searrow$}
 \rput(1.5,0){$0$}
 \rput(2.2,0.7){$\swarrow$}
 \rput(2.95,1.5){$1$}
 \rput(3.65,0.7){$\searrow$}
 \rput(4.37,0){$0$}
 \rput(5.1,0.7){$\swarrow$}
 \rput(5.85,1.5){$1$}
\end{pspicture} 
\\[.3cm]
\ee
By construction, the $\Wc$-extended rank-2 modules are all self-contragredient:
\be
 (\Rh_r^b)^\ast=\Rh_r^b
\ee

To describe the fusion algebra 
\be
 \big\langle\ketw{r,s},\ketwa{r,s};\,r\in\mathbb{Z}_{1,2},\,s\in\mathbb{N}\big\rangle
\ee
generated by repeated fusion of the
$\Wc$-Kac representations and their contragredient counterparts,
we mimic (\ref{C}) and introduce
\be
 \Cch_n[\ketw{r,s}]=\begin{cases} 
  \ketw{r,s},\quad &n>0
  \\[.2cm]
  \ketwa{r,s},\quad &n<0
 \end{cases}
\label{Ch}
\ee
In our applications, $\Cch_0[\ketw{r,s}]$ only appears if $\ketw{r,s}$ is fully reducible
in which case
\be
 \Cch_0[\ketw{r,s}]=\ketw{r,s}=\ketwa{r,s},\qquad s\in\mathbb{Z}_{1,p-1}\cup p\mathbb{N}
\ee
The fusion rules involving the contragredient modules $\ketwa{r,s}$
are inferred from the corresponding fusion rules in the Virasoro picture~\cite{Ras1012}.
We thus find that the decomposition of the fusion products 
\be
 \ketwa{r,s}\fus\ketwa{r',s'}=\big(\ketw{r,s}\fus\ketw{r',s'}\big)^\ast,\qquad
 \Rh_r^b\fus\ketwa{r',s'}=\Rh_r^b\fus\ketw{r',s'}
\label{Rketwa}
\ee
follow readily from the fusion rules underlying the
$\Wc$-Kac fusion algebra discussed above, while
\bea
 \ketw{r,b+kp}\fus\ketwa{r',b'+k'p}
  &=&kk'\Big(\bigoplus_\beta^{p-b-b'-1}\Rh_{r\cdot r'\cdot k\cdot k'}^\beta\Big)
  \oplus(k+1)(k'+1)\Big(\bigoplus_\beta^{b+b'-p-1}\Rh_{r\cdot r'\cdot k\cdot k'}^\beta\Big)\nn
 &\oplus&\big(kk'+\min(k,k')\big)\Big(\bigoplus_\beta^{p-|p-b-b'|-1}
   \Rh_{2\cdot r\cdot r'\cdot k\cdot k'}^\beta\Big)\nn
 &\oplus&|k-k'|\Big(\bigoplus_\beta^{(b-b')\mathrm{sg}(k'-k)-1}
    \Rh_{2\cdot r\cdot r'\cdot k\cdot k'}^\beta\Big)\nn
 &\oplus&\bigoplus_{\beta=|b-b'|+1,\,\mathrm{by}\,2}^{p-|p-b-b'|-1}
  \Cch_{k-k'}[\ketw{r\cdot r',\beta+|k-k'|p}]
\label{bkbkast}
\eea
The set of distinct, $\Wc$-indecomposable representations partaking in this 
contragrediently-extended $\Wc$-Kac fusion algebra is
\be
 \Jc_\Wc^\mathrm{Cont}
 =\Jc^\mathrm{Kac}_\Wc\cup\{\ketwa{r,s};\,r\in\mathbb{Z}_{1,2},\,
 s\in\mathbb{N}\setminus\big(\mathbb{Z}_{1,p-1}\cup p\mathbb{N}\big)\}
\ee
here written as a disjoint union of sets.
It is noted that the fusion subalgebras generated by the $\Wc$-Kac representations
and their contragredient counterparts, respectively, are isomorphic
\be
 \big\langle\ketw{r,s};\,r\in\mathbb{Z}_{1,2},\,s\in\mathbb{N}\big\rangle\,\simeq\,
 \big\langle\ketwa{r,s};\,r\in\mathbb{Z}_{1,2},\,s\in\mathbb{N}\big\rangle
\label{KacContKac}
\ee 
This resembles the situation (\ref{KacKac}) in the Virasoro picture.

Combining the second fusion property in (\ref{Rketwa}) with
the fact that the $\Wc$-projective modules form an ideal of the $\Wc$-Kac 
fusion algebra, we see that the $\Wc$-projective modules also form an ideal
of the larger contragrediently-extended $\Wc$-Kac fusion algebra 
$\langle\Jc_\Wc^\mathrm{Cont}\rangle$. 
It also follows that the insensitivity properties (\ref{Winsensitive})
of the $\Wc$-projective modules are supplemented by
\be
 \Rh_r^b\fus\ketwa{r',b'+k'p}
   =\Rh_r^b\fus\big[k'\ketw{r'\cdot k',p-b'}\oplus(k'+1)\ketw{2\cdot r'\cdot k',b'}\big]
\ee
implying that 
fusion by $\Rh_r^b$ is an {\em exact functor} in $\langle\Jc_\Wc^\mathrm{Cont}\rangle$
for all $\Rh_r^b\in\Jc^\mathrm{Proj}_\Wc$.

Following up on the discussion at the end of Section~\ref{SecFusVir},
the lattice issue with Kac representations versus contragredient Kac representations carries
over to the $\Wc$-extended picture. Here we have merely reiterated the assertion~\cite{Ras1012}
that it is the Kac representations $(r,s)$ and not (in general) the contragredient Kac 
representations $(r,s)^\ast$
which are associated with Yang-Baxter integrable boundary conditions in the lattice approach.
As a consequence of the way we have introduced the corresponding Yang-Baxter
integrable boundary conditions in the $\Wc$-extended picture, it is the
$\Wc$-Kac representations $\ketw{r,s}$ and not (in general) their contragredient
counterparts $\ketwa{r,s}$ which are associated with $\Wc$-extended boundary conditions.
The two families generate isomorphic fusion algebras (\ref{KacContKac})
and the corresponding characters are identical $\chih_{r,s}(q)=\chih^\ast_{r,s}(q)$.

\section{Polynomial fusion rings}
\label{SecPol}

Our last objectives are to determine polynomial fusion rings isomorphic with the 
$\Wc$-Kac fusion algebra $\langle\Jc_\Wc^\mathrm{Kac}\rangle$
and its contragredient extension $\langle\Jc_\Wc^\mathrm{Cont}\rangle$, and to
identify the corresponding Grothendieck ring of characters.
This is a continuation of our recent work~\cite{Ras0812,Ras0906,Ras0908} 
on polynomial fusion rings in ${\cal WLM}(p,p')$.
The corresponding constructions in the Virasoro picture were obtained
in~\cite{RP0709}.

\subsection{$\Wc$-extended Kac fusion algebra}

Together with the fact that the $\Wc$-extended fundamental fusion algebra 
$\langle\Jc_\Wc^\mathrm{Fund}\rangle$ is a subalgebra of the $\Wc$-Kac fusion algebra, 
the fusion rules
\bea
 \ketw{1,2}\fus\ketw{1,b+kp}&=&\ketw{1,b-1+kp}\oplus\ketw{1,b+1+kp}\nn
 \ketw{1,p+1}\fus\ketw{1,b+kp}&=&k\Big(\bigoplus_{\beta}^{p-b}\Rh_{1\cdot k}^\beta\Big)
  \oplus(k+1)\Big(\bigoplus_\beta^{b-2}\Rh_{2\cdot k}^\beta\Big)
  \oplus\ketw{1,b+(k+1)p}
\label{W12p}
\eea
where $b\in\mathbb{Z}_{1,p-1}$,
demonstrate that the $\Wc$-Kac fusion algebra
is generated from repeated fusion of the four modules
$\ketw{1,1}$, $\ketw{2,1}$, $\ketw{1,2}$ and $\ketw{1,p+1}$,
that is,
\be
 \big\langle\Jc_\Wc^{\mathrm{Kac}}\big\rangle
 =\big\langle\ketw{1,1},\ketw{2,1},\ketw{1,2},\ketw{1,p+1}\big\rangle
\label{WKac4}
\ee
Since $\ketw{1,1}=\Mh_{1,1}$ is the algebra identity,
it is therefore natural to expect that this fusion algebra is
isomorphic to a polynomial ring in the three entities
$X\leftrightarrow\ketw{2,1}=\Mh_{2,1}$, $Y\leftrightarrow\ketw{1,2}=\Mh_{1,2}$ 
and $Z\leftrightarrow\ketw{1,p+1}$. 
This is indeed what we find and it is the content of Proposition 1 below.
In the following, $T_n(x)$ and $U_n(x)$ are Chebyshev polynomials of the first and second kind, 
respectively, where we set $U_{-1}(x)=0$.
\\[.2cm]
{\bf Proposition 1.}\quad The $\Wc$-Kac fusion algebra is isomorphic to the polynomial ring 
generated by $X$, $Y$ and $Z$ modulo the ideal
\be
 \Ic_\Wc^\mathrm{Kac}=(X^2-1,P_p(X,Y),Q_p(Y,Z))
\ee
that is,
\be
 \big\langle\Jc_\Wc^{\mathrm{Kac}}\big\rangle
  \simeq\mathbb{C}[X,Y,Z]/\Ic_\Wc^\mathrm{Kac}
\ee
where
\be
 P_p(X,Y)=\big[X-T_p(\tfrac{Y}{2})\big]U_{p-1}(\tfrac{Y}{2}),\qquad
 Q_p(Y,Z)=\big[Z-U_p(\tfrac{Y}{2})\big]U_{p-1}(\tfrac{Y}{2})
\label{PQ}
\ee
For $r\in\mathbb{Z}_{1,2}$, $b\in\mathbb{Z}_{0,p-1}$ and $k\in\mathbb{N}_0$, 
the isomorphism reads
\bea
 \ketw{r,b+kp}&\leftrightarrow& 
   X^{r-1}
    \Big(U_{kp+b-1}(\tfrac{Y}{2})+\big[Z^k-U_{p}^k(\tfrac{Y}{2})\big]U_{b-1}(\tfrac{Y}{2})\Big)\nn
 \Rh_r^b&\leftrightarrow&
   (2-\delta_{b,0})X^{r-1}T_b(\tfrac{Y}{2})U_{p-1}(\tfrac{Y}{2})
\eea
{\bf Proof.}\quad The relation $P_p(X,Y)=0$ corresponds to the identification
$\ketw{1,2p}\equiv2\ketw{2,p}$, 
while the relation $Q_p(Y,Z)=0$ corresponds to the fusion rule
\be
 \ketw{1,p}\fus\ketw{1,p+1}=2\ketw{2,p}\oplus\bigoplus_{\beta}^{p-2}\Rh_1^\beta
\ee
when employing the identity
\be
 \sum_{\beta=\eps(p),\,\mathrm{by}\,2}^p\big(2-\delta_{\beta,0}\big)T_\beta(x)=U_p(x)
\ee
The remaining fusion rules are then verified straightforwardly in the polynomial ring.
Here we only demonstrate explicitly the two fusion rules in (\ref{W12p}).
The first of these follows immediately from the recursion relation for the Chebyshev
polynomials. To show the second of the fusion rules, we follow~\cite{Ras1012} on
the similar fusion rule in the Virasoro picture and note
the basic decomposition rules
\be
 U_m(x)U_n(x)=\sum_{j=|m-n|,\,\mathrm{by}\,2}^{m+n}U_j(x),\qquad
 2T_m(x)U_{n-1}(x)=U_{n+m-1}(x)+\mathrm{sg}(n-m)U_{|n-m|-1}(x)
\ee
As a consequence, we have
\be
 U_{p-1}(x)\sum_{j=0}^{k-1}U_p^{k-j-1}(x)U_{jp+b-2}(x)=U_{b-1}(x)U_p^k(x)-U_{kp+b-1}(x)
\ee
which is established by induction in $k$ and shows that the expression on the right side is
divisible by $U_{p-1}(x)$. This is of importance when multiplied by $Z$ due to
the form of $Q_p(Y,Z)$. With the additional observation that
\be
 U_{r-1}(\tfrac{X}{2})U_{p-1}(\tfrac{Y}{2})\equiv U_{rp-1}(\tfrac{Y}{2})\quad 
   (\mathrm{mod}\ P_p(X,Y))
\label{UU}
\ee
which follows by induction in $r$, the second fusion rule readily follows.
$\quad\Box$
\\[.2cm]
In~\cite{Ras1012}, we demonstrated that the conjectured Kac fusion algebra
$\langle\Jc^\mathrm{Kac}\rangle$ in the
Virasoro picture is isomorphic to the polynomial ring
\be
 \big\langle\Jc^{\mathrm{Kac}}\big\rangle\simeq C[X,Y,Z]/(P_p(X,Y),Q_p(Y,Z))
\label{JKacC}
\ee
In somewhat sloppy notation, we thus have the relation
\be
 \big\langle\Jc_\Wc^{\mathrm{Kac}}\big\rangle
  \simeq\big\langle\Jc^{\mathrm{Kac}}\big\rangle/(X^2-1)
\label{JJ}
\ee
between the $\Wc$-Kac fusion algebra and the Kac fusion
algebra itself.

\subsection{Contragredient extension}

Extending the arguments presented above for the $\Wc$-Kac fusion algebra, one
finds that its contragredient extension $\langle\Jc_\Wc^\mathrm{Cont}\rangle$
is also generated from repeated fusion of a small number of modules, namely
\be
 \big\langle\Jc_\Wc^\mathrm{Cont}\big\rangle
 =\big\langle\Mh_{1,1},\Mh_{2,1},\Mh_{1,2},\ketw{1,p+1},\ketwa{1,p+1}\big\rangle
\label{WKac5}
\ee
where it is recalled that
\be
 \Mh_{1,1}=\ketw{1,1}=\ketwa{1,1},\qquad
 \Mh_{2,1}=\ketw{2,1}=\ketwa{2,1},\qquad
 \Mh_{1,2}=\ketw{1,2}=\ketwa{1,2}
\ee
Using
\be
 \ketw{1,p+1}\fus\ketwa{1,p+1}=\Mh_{1,1}
  \oplus2\Rh_2^1
  \oplus\bigoplus_\beta^{p-3}\Rh_1^\beta
\label{1p1p}
\ee
in particular,
one also finds that $\langle\Jc_\Wc^\mathrm{Cont}\rangle$ is isomorphic to a polynomial ring
in the four entities
$X\leftrightarrow\Mh_{2,1}$, $Y\leftrightarrow\Mh_{1,2}$, $Z\leftrightarrow\ketw{1,p+1}$
and $Z^\ast\leftrightarrow\ketwa{1,p+1}$ as demonstrated in Proposition 2 below.
\\[.2cm]
\noindent
{\bf Proposition 2.}\quad The contragrediently-extended $\Wc$-Kac fusion algebra
is isomorphic to the polynomial ring generated
by $X$, $Y$, $Z$ and $Z^\ast$ modulo the ideal 
\be
 \Ic_\Wc^\mathrm{Cont}=(X^2-1,P_p(X,Y),Q_p(Y,Z),Q_p(Y,Z^\ast),R_p(Y,Z,Z^\ast))
\ee 
that is,
\be
 \big\langle\Jc_\Wc^{\mathrm{Cont}}\big\rangle
  \simeq\mathbb{C}[X,Y,Z,Z^\ast]/\Ic_\Wc^\mathrm{Cont}
\ee
where the polynomials $P_p$ and $Q_p$ are defined in (\ref{PQ}) while
\be
 R_p(Y,Z,Z^\ast)=ZZ^\ast-U_p^2(\tfrac{Y}{2})
\label{R}
\ee
For $r\in\mathbb{Z}_{1,2}$, $b\in\mathbb{Z}_{0,p-1}$ and $k\in\mathbb{N}_0$, 
the isomorphism reads
\bea
 \ketw{r,b+kp}&\leftrightarrow& 
   X^{r-1}
    \Big(U_{kp+b-1}(\tfrac{Y}{2})+\big[Z^k-U_{p}^k(\tfrac{Y}{2})\big]U_{b-1}(\tfrac{Y}{2})\Big)\nn
 \ketwa{r,b+kp}&\leftrightarrow& 
   X^{r-1}
    \Big(U_{kp+b-1}(\tfrac{Y}{2})+\big[(Z^\ast)^k-U_{p}^k(\tfrac{Y}{2})\big]U_{b-1}(\tfrac{Y}{2})\Big)\nn
 \Rh_r^b&\leftrightarrow&
   (2-\delta_{b,0})X^{r-1}T_b(\tfrac{Y}{2})U_{p-1}(\tfrac{Y}{2})
\eea
{\bf Proof.}\quad This proof is almost identical to the proof in~\cite{Ras1012} of the similar
proposition in the Virasoro picture, but is included for completeness. 
Compared to the proof of Proposition 1, the essential new feature
is the appearance of $Z^\ast$. The relation $Q_p(Y,Z^\ast)=0$ plays the same
role for the contragredient $\Wc$-Kac representations and $Z^\ast$ as
$Q_p(Y,Z)=0$ does for the $\Wc$-Kac representations and $Z$. This yields the
part of the polynomial ring corresponding to (\ref{KacContKac}).
The relation $R_p(Y,Z,Z^\ast)=0$ corresponds to the fusion rule (\ref{1p1p}).
To establish the general fusion rule (\ref{bkbkast}) in the ring picture, we first use induction in $n$
to establish
\be
 U_p^{2n}(\tfrac{Y}{2})Z^m\equiv Z^m
   +\sum_{j=0}^{n-1}U_p^{m+2j}(\tfrac{Y}{2})U_{p-1}(\tfrac{Y}{2})
   U_{p+1}(\tfrac{Y}{2})
   \quad (\mathrm{mod}\ Q_p(Y,Z)),\qquad n\in\mathbb{N}
\ee
and similarly for $Z$ replaced by $Z^\ast$. This is needed when reducing
\be
 Z^k(Z^\ast)^{k'}\equiv U_p^{2\min(k,k')}(\tfrac{Y}{2})\begin{cases} Z^{k-k'},\quad&k\geq k' \\ 
   (Z^\ast)^{k'-k},\quad&k<k'  \end{cases}
   \qquad  (\mathrm{mod}\ R_p(Y,Z,Z^\ast))
\ee
For simplicity, we let $k\geq k'$ in which case we find
\be
 \ketw{1,b+kp}\fus\ketwa{1,b'+k'p}\leftrightarrow
  \big[Z^{k-k'}-U_p^{k-k'}(\tfrac{Y}{2})\big]U_{b-1}(\tfrac{Y}{2})U_{b'-1}(\tfrac{Y}{2})
    +U_{kp+b-1}(\tfrac{Y}{2})U_{k'p+b'-1}(\tfrac{Y}{2})
\ee
This polynomial expression is recognized as corresponding to the right side of (\ref{bkbkast}).
$\quad\Box$
\\[.2cm]
In~\cite{Ras1012}, we demonstrated that the conjectured contragrediently-extended
Kac fusion algebra $\langle\Jc^\mathrm{Cont}\rangle$ in the
Virasoro picture is isomorphic to the polynomial ring
\be
 \big\langle\Jc^{\mathrm{Cont}}\big\rangle\simeq 
   C[X,Y,Z,Z^\ast]/(P_p(X,Y),Q_p(Y,Z),Q_p(Y,Z^\ast),R_p(Y,Z,Z^\ast))
\label{JContC}
\ee
In somewhat sloppy notation, we thus have the relation
\be
 \big\langle\Jc_\Wc^{\mathrm{Cont}}\big\rangle
  \simeq\big\langle\Jc^{\mathrm{Cont}}\big\rangle/(X^2-1)
\ee
between the contragrediently-extended $\Wc$-Kac fusion algebra and the 
contragrediently-extended Kac fusion algebra itself.
With this and (\ref{JJ}) in mind, the proofs of Proposition 1 and 2 could have been
reduced to an analysis of the consequences of $X^2=1$ since (\ref{JKacC}) and
(\ref{JContC}) were established in~\cite{Ras1012}. However, we found it more
instructive to include direct and independent proofs of the two propositions above.

\subsection{Grothendieck ring}

The set of Virasoro characters in a CFT naturally forms 
a Grothendieck group whose generators are equivalence classes $[R]$ formed
by the characters: $[R]=\chit[R](q)$. 
Its group operation is addition and is defined via direct 
summation of the representations of the equivalence classes
\be
 [R_1]+[R_2]=[R_1\oplus R_2]
\ee
that is, by addition of characters.
For rational CFTs, this Grothendieck group admits a ring structure whose multiplication 
follows from the fusion product of representations
\be
 [R_1]\ast[R_2]=[R_1\otimes R_2]
\label{R1R2}
\ee
For logarithmic models, on the other hand, the fusion of representations does not, in 
general, induce a product on the Grothendieck group in this way, see \cite{GRW0905} for 
example.
However, on the Grothendieck group associated
with the fundamental fusion algebra of ${\cal WLM}(1,p)$, the fusion rules {\em do}
induce a well-defined multiplication (\ref{R1R2})
thereby turning the group into a ring, as described in~\cite{PRR0907}.

The Grothendieck group associated with the fundamental fusion algebra
of ${\cal WLM}(1,p)$ is generated by the $2p$ generators
\be
 G_{r,s}=[\Mh_{r,s}],\qquad r\in\mathbb{Z}_{1,2},\ s\in\mathbb{Z}_{1,p}
\label{G}
\ee
corresponding to the set of irreducible modules.
Following from (\ref{Rdiagram}), the equivalence class of a rank-2 module thus decomposes as
\be
  [\Rh_{r}^{b}]= 
  2G_{2\cdot r,b}+2G_{1\cdot r.p-b},\qquad b\in\mathbb{Z}_{1,p-1}
\label{RGG}
\ee
The corresponding multiplication rules follow from the fusion rules and are given by
\be
  G_{r,s}\ast G_{r',s'}\ =\ 
  \!\sum_{j=|s-s'|+1,\ \!\mathrm{by}\ \!2}^{p-|p-s-s'|-1}
  \!\!\! G_{r\cdot r',j}
    +\!\!\sum_{\beta=\eps(s+s'-p-1),\ \!\mathrm{by}\ \!2}^{s+s'-p-1}
    \!\!\!(2-\delta_{\beta,0})\big(G_{r\cdot r',p-\beta}+G_{2\cdot r\cdot r',\beta}\big)   
\label{Gmult}
\ee
where $G_{r,0}\equiv0$.

Let
\be
   \big\langle \Jc_\Wc^\mathrm{Grot}\big\rangle,\qquad 
    \Jc_\Wc^\mathrm{Grot}=\{G_{r,s};\ r\in\mathbb{Z}_{1,2},
     s\in\mathbb{Z}_{1,p}\}
\ee
denote the Grothendieck ring associated with ${\cal WLM}(1,p)$, and let $\sim$ denote the
equivalence relation between modules in $\Jc^\mathrm{Cont}_\Wc$
with identical characters such that
\be
  \Rh_{r}^{b}
  \,\sim\,2\Mh_{2\cdot r,b}\oplus2\Mh_{1\cdot r,p-b},\qquad
 \ketw{r,s_0+kp}\,\sim\,\ketwa{r,s_0+kp}
  \,\sim\,
  k\Mh_{r\cdot k,p-s_0}\oplus(k+1)\Mh_{2\cdot r\cdot k,s_0}
\label{RMM}
\ee
%
\noindent
{\bf Proposition 3.}\quad The Grothendieck ring associated with ${\cal WLM}(1,p)$
is isomorphic with the contragredient, the $\Wc$-Kac and the fundamental 
fusion algebra modulo the equivalence relation $\sim$,
that is,
\be
 \big\langle \Jc_\Wc^\mathrm{Grot}\big\rangle
  \simeq
   \Big(\big\langle \Jc_\Wc^\mathrm{Cont}\big\rangle\big/\!\sim\Big)
  \simeq
   \Big(\big\langle \Jc_\Wc^\mathrm{Kac}\big\rangle\big/\!\sim\Big)
  \simeq
   \Big(\big\langle \Jc_\Wc^\mathrm{Fund}\big\rangle\big/\!\sim\Big)
\label{GCKF}
\ee
It is also isomorphic with the following polynomial rings
\be
 \big\langle \Jc_\Wc^\mathrm{Grot}\big\rangle
  \simeq
 \mathbb{C}[X,Y]/\big(X^2-1,X-T_p(\tfrac{Y}{2})\big),\qquad
 \big\langle \Jc_\Wc^\mathrm{Grot}\big\rangle
  \simeq
  \mathbb{C}[Y]/\big((Y^2-4)U_{p-1}^2(\tfrac{Y}{2})\big)
\label{Grot}
\ee
The isomorphisms in (\ref{Grot}) are given by
\be
  G_{r,s}\ \leftrightarrow\ X^{r-1}U_{s-1}(\tfrac{Y}{2}),\qquad
 G_{r,s}\ \leftrightarrow\ T_p^{r-1}(\tfrac{Y}{2})U_{s-1}(\tfrac{Y}{2}),\qquad
  r\in\mathbb{Z}_{1,2},\ s\in\mathbb{Z}_{1,p}
\ee
respectively.
\\[.2cm]
{\bf Proof.}\quad That the Grothendieck ring is isomorphic with 
$\langle \Jc_\Wc^\mathrm{Fund}\rangle/\!\sim$ was established in~\cite{PRR0907}.
The other two isomorphisms in (\ref{GCKF}) correspond to an elevation of this
result to the $\Wc$-Kac fusion algebra and further to
the contragredient extension thereof. These elevations are established by applying
\be
 [\ketw{r,s_0+kp}]=[\ketwa{r,s_0+kp}]=
  kG_{r\cdot k,p-s_0}+(k+1)G_{2\cdot r\cdot k,s_0}
\label{rsG}
\ee
and the multiplication rule (\ref{Gmult}) to the fusion rules (\ref{rbrb}), (\ref{fusRKac}) 
and (\ref{bkbkast}) involving $\ketw{r,s_0+kp}$ or $\ketwa{r,s_0+kp}$, where
(\ref{rsG}) itself is a consequence of (\ref{RMM}).

To establish the first isomorphism in (\ref{Grot}), we note that 
$\Rh_1^1\sim2\Mh_{2,1}\oplus2\Mh_{1,p-1}$ implies the equivalence relation
\be
 X\,\sim\,T_1(\tfrac{Y}{2})U_{p-1}(\tfrac{Y}{2})-U_{p-2}(\tfrac{Y}{2})
 =T_p(\tfrac{Y}{2})
\ee
As a consequence, 
$\ketw{1,p+1}\sim\Mh_{1,p-1}\oplus\Mh_{2,1}$ implies the equivalence relation
\be
 Z\,\sim\,U_{p-2}(\tfrac{Y}{2})+2X\,\sim\,U_{p-2}(\tfrac{Y}{2})+2T_p(\tfrac{Y}{2})
  =U_p(\tfrac{Y}{2})
\ee
We likewise have $Z^\ast\sim U_p(\tfrac{Y}{2})$.
The polynomials $X-T_p(\tfrac{Y}{2})$,
$Z-U_p(\tfrac{Y}{2})$ and $Z^\ast-U_p(\tfrac{Y}{2})$ are divisors of the polynomials 
$P_p(X,Y)$, $Q_p(Y,Z)$ and $Q_p(Y,Z^\ast)$, respectively,
as defined in (\ref{PQ}), and since $R_p(Y,Z,Z^\ast)$ is trivial modulo 
$Z-U_p(\tfrac{Y}{2}),Z^\ast-U_p(\tfrac{Y}{2})$, 
they eliminate the dependence on $Z$ and $Z^\ast$ in the 
polynomial ring in Proposition 2. They also simplify the polynomials associated with the
(contragredient) $\Wc$-Kac representations as we have
\bea
 \ketw{r,b+kp}&\sim&\ketwa{r,b+kp}\,\sim\,X^{r-1}U_{kp+b-1}(\tfrac{Y}{2})\nn
 &\equiv&kX^{r\cdot k-1}U_{p-b-1}(\tfrac{Y}{2})
   +(k+1)X^{2\cdot r\cdot k-1}U_{b-1}(\tfrac{Y}{2})
   \qquad (\mathrm{mod}\ X^2-1,X-T_p(\tfrac{Y}{2}))\nn
 &\sim&k\Mh_{r\cdot k,p-b}\oplus(k+1)\Mh_{2\cdot r\cdot k,b}
\eea
where the polynomial equivalence follows by induction in $k$. 
Likewise, the polynomial realizations of the rank-2 modules simplify as
\bea
   \Rh_{r}^{b}
    &\sim&2X^{r-1}T_b(\tfrac{Y}{2})U_{p-1}(\tfrac{Y}{2})
  \equiv2X^{2\cdot r-1}U_{b-1}(\tfrac{Y}{2})+2X^{r-1}U_{p-b-1}(\tfrac{Y}{2})
   \qquad (\mathrm{mod}\ X-T_p(\tfrac{Y}{2}))\nn
  &\sim&2\Mh_{2\cdot r,b}\oplus2\Mh_{1\cdot r,p-b}
\eea
This completes the proof of the first isomorphism in (\ref{Grot}).
The reduction of the polynomial ring in the two variables $X$ and $Y$ in (\ref{Grot})
to the polynomial ring in the single variable $Y$ follows from
\be
 X^2-1\equiv T_p^2(\tfrac{Y}{2})-1=\tfrac{1}{4}(Y^2-4)U_{p-1}^2(\tfrac{Y}{2})
 \qquad (\mathrm{mod}\ X-T_p(\tfrac{Y}{2}))
\ee
$\Box$
\\[.2cm]
As illustration of the structure of the Grothendieck rings, we follow~\cite{PRR0907} and
consider ${\cal WLM}(1,2)$ whose four-dimensional Grothendieck ring  
\be
 \big\langle \Jc_\Wc^\mathrm{Grot}\big\rangle
  \simeq\mathbb{C}[Y]/(Y^4-4Y^2)
\label{Grot2}
\ee
is generated by
\be
 G_{1,1}\ \leftrightarrow\ 1,\qquad
 G_{1,2}\ \leftrightarrow\ Y,\qquad
 G_{2,1}\ \leftrightarrow\ \tfrac{1}{2}Y^2-1,\qquad
 G_{2,2}\ \leftrightarrow\ \tfrac{1}{2}Y^3-Y
\label{Gks2}
\ee
The multiplication rules are given in the Cayley tables in Figure~\ref{Cayley}.
\psset{unit=1cm}
\begin{figure}
$$
\renewcommand{\arraystretch}{1.5}
\begin{array}{c||cccc}
\ast&G_{1,1}&G_{2,1}&G_{1,2}&G_{2,2}\\[4pt]
\hline \hline
\rule{0pt}{14pt}
 G_{1,1}&G_{1,1}&G_{2,1}&G_{1,2}&G_{2,2}
    \\[4pt]
 G_{2,1}&G_{2,1}&G_{1,1}&G_{2,2}&G_{1,2}
    \\[4pt]
 G_{1,2}&G_{1,2}&G_{2,2}&2G_{1,1}+2G_{2,1}&2G_{1,1}+2G_{2,1}
    \\[4pt]
 G_{2,2}&G_{2,2}&G_{1,2}&2G_{1,1}+2G_{2,1}&2G_{1,1}+2G_{2,1}
\end{array}
\hspace{1.3cm}
\begin{array}{c||cccc}
\ast&0&1&-\tfrac{1}{8}&\tfrac{3}{8}\\[4pt]
\hline \hline
\rule{0pt}{14pt}
 0&0&1&-\tfrac{1}{8}&\tfrac{3}{8}
    \\[4pt]
 1&1&0&\tfrac{3}{8}&-\tfrac{1}{8}
    \\[4pt]
 -\tfrac{1}{8}&-\tfrac{1}{8}&\tfrac{3}{8}&2(0)+2(1)&2(0)+2(1)
    \\[4pt]
 \tfrac{3}{8}&\tfrac{3}{8}&-\tfrac{1}{8}&2(0)+2(1)&2(0)+2(1)
\end{array}
$$
\caption{Cayley tables of the multiplication rules for $\langle \Jc_\Wc^\mathrm{Grot}\rangle$
in ${\cal WLM}(1,2)$. 
In the second table, the generators $G_{r,s}$ are represented by the conformal
weights of the corresponding $\Wc$-irreducible modules.}
\label{Cayley}
\end{figure}

\section{Discussion}
\label{SecDisc}

We have constructed new Yang-Baxter integrable boundary conditions giving
rise to reducible yet indecomposable rank-1 representations in the $\Wc$-extended
logarithmic minimal model ${\cal WLM}(1,p)$ where $p=2,3,\ldots$. 
These $\Wc$-Kac representations $\ketw{r,s}$
correspond to finitely-generated $\Wc$-extended Feigin-Fuchs modules over the
$W$-algebra $\Wc(p)$, and their fusion properties were inferred from the
fusion rules in the Virasoro picture ${\cal LM}(1,p)$ of the logarithmic minimal model.
The contragredient modules $\ketwa{r,s}$ to the $\Wc$-Kac representations were also
introduced, and the correspondingly-extended fusion algebra was derived.
Polynomial fusion rings isomorphic with the various fusion algebras were
subsequently determined, and the corresponding Grothendieck ring 
of characters was identified.

The results presented here pertain to the $\Wc$-extended
logarithmic minimal models ${\cal WLM}(1,p)$
and are based on the work~\cite{Ras1012} on the same models in the
Virasoro picture ${\cal LM}(1,p)$.
The methods used to obtain the various results, on the other hand, 
are expected to be applicable also in the general cases ${\cal LM}(p,p')$ and ${\cal WLM}(p,p')$,
at least after implementation of the disentangling procedure employed in~\cite{Ras0805}
when extending the work~\cite{PRR0803} on ${\cal WLM}(1,p)$ to ${\cal WLM}(p,p')$.
We hope to discuss these generalizations elsewhere, in particular for
critical percolation as described by ${\cal LM}(2,3)$ and ${\cal WLM}(2,3)$.

As already mentioned, the category of $\Wc(p)$-modules and
the category of finite-dimensional $\bar{U}_q(sl_2)$-modules  at $q=e^{\pi i/p}$
are equivalent as abelian categories for all 
$p\geq2$~\cite{FGST0504,FGST0512,FGST0606hep,FGST0606math,NT0902}.
For $p\geq3$, however, it was found~\cite{KS0901}
that they are not equivalent as braided tensor categories 
due to complications arising from the presence of the modules 
${\cal E}_s^\pm(n;\lambda)$.
These `circular' modules were denoted by ${\cal O}_s^\pm(n,z)$ in~\cite{FGST0512} 
where they first appeared.
We note that the subcategory of $\bar{U}_q(sl_2)$-modules obtained by excluding these 
circular modules closes under tensor products. Likewise, we can define the 
`contragrediently-extended boundary $\Wc(p)$-category' 
as the subcategory of $\Wc(p)$-modules associated with the ${\cal WLM}(1,p)$
boundary conditions constructed in~\cite{PRR0803} and 
in Section~\ref{SecW}, supplemented by the $\Wc$-reducible yet $\Wc$-indecomposable 
contragredient $\Wc$-Kac 
representations (thus counting the $\Wc$-irreducible representations only once, cf. (\ref{KastK})). 
Our results then suggest that this contragrediently-extended
boundary $\Wc(p)$-category and the above subcategory of $\bar{U}_q(sl_2)$-modules
{\em are} equivalent as tensor categories. That is, these proposed subcategories are not
only equivalent as abelian categories, but we have verified that their tensor
structures are compatible. To facilitate this verification, we refer to the dictionary 
in Appendix~\ref{AppDic}. Without extending the category of boundary $\Wc(p)$-modules by the 
contragredient $\Wc$-Kac representations,
the boundary category {\em itself} is equivalent to the corresponding 
subcategory of $\bar{U}_q(sl_2)$-modules as tensor categories.
These affirmative observations provide substantial evidence
for the conjectured Kac fusion algebra of \cite{Ras1012} and its elevation to the
$\Wc$-extended picture discussed in the present work. They also support the
Kazhdan-Lusztig dualities of~\cite{FGST0504,FGST0512,FGST0606hep,FGST0606math}
and~\cite{BFGT0901,BGT1102}.

The recent works~\cite{GRW1008,PR1010} on the structure of bulk logarithmic CFTs and 
their relation
with boundary logarithmic CFTs have greatly advanced our understanding of logarithmic CFT.
However, these results are based on the {\em `rational'} part of the $\Wc$-extended picture formed
by the {\em finitely} many $\Wc$-irreducible and $\Wc$-projective modules on the boundary side.
The present work suggests a much richer boundary model, even in the logarithmic minimal model
${\cal WLM}(1,p)$, and it would be interesting to readdress the relation
with the bulk model in light of these new findings.
\section*{Acknowledgments}
\vskip.1cm
\noindent
This work is supported by the Australian Research Council under the Future Fellowship
scheme, project number FT100100774.
The author thanks Paul A. Pearce and Ilya Yu.\! Tipunin
for helpful discussions and comments.

\appendix

\section{Fusion rules in ${\cal LM}(1,p)$}
\label{AppFusVir}

\subsection{Kac fusion rules}
\label{AppKacRules}

{}From~\cite{RP0707,Ras1012}, for $b,b'\in\mathbb{Z}_{0,p-1}$ and $k,k'\in\mathbb{N}_0$,
we have the fusion rules
\bea
 (1,b+kp)\otimes(1,b'+k'p)&=&\bigoplus_{j=|k-k'|+1,\,\mathrm{by}\,2}^{k+k'-1}
   \!\!\bigoplus_{\beta}^{p-|b-b'|-1}\R_{j}^\beta
   \oplus\bigoplus_{j=|k-k'+\mathrm{sg}(b-b')|+1,\,\mathrm{by}\,2}^{k+k'}
   \!\!\bigoplus_{\beta}^{|b-b'|-1}\R_{j}^\beta\nn
 &\oplus&\bigoplus_{\beta}^{b+b'-p-1}\R_{k+k'+1}^\beta
   \oplus\bigoplus_{\beta=|b-b'|+1,\,\mathrm{by}\,2}^{p-|p-b-b'|-1}(1,\beta+(k+k')p)\nn
 \R_{1}^{b}\otimes(1,b'+k'p)&=&\bigg(
  \bigoplus_{\beta}^{p-|b-b'|-1}\R_{k'}^\beta
  \oplus
  \!\!\bigoplus_{\beta}^{p-b-b'-1}\R_{k'}^\beta
  \oplus(1-\delta_{k',0})\!\!\bigoplus_{\beta}^{b-b'-1}(\R_{k'-1}^\beta\oplus\R_{k'+1}^\beta)
  \nn
 &\oplus& \bigoplus_{\beta=|b-b'|+1,\,\mathrm{by}\,2}^{p-|p-b-b'|-1}\R_{k'+1}^\beta
  \oplus2\Big(\bigoplus_{\beta}^{b'-b-1}\R_{k'+1}^\beta\Big)
  \oplus\!\!\bigoplus_{\beta}^{b+b'-p-1}\R_{k'+2}^\beta 
  \bigg)/(1+\delta_{b,0})\nn
  \R_1^b\otimes\R_{1}^{b'}&=&\bigg(
  \bigoplus_{\beta=|p-b-b'|+1,\,\mathrm{by}\,2}^{p-|b-b'|-1}\R_{1}^\beta
  \oplus\!\!\bigoplus_{\beta}^{p-|b-b'|-1}\R_1^\beta
  \oplus3\Big(\bigoplus_{\beta}^{p-b-b'-1}\R_{1}^\beta\Big)
  \oplus\!\bigoplus_{\beta}^{|b-b'|-1}\R_2^\beta
  \nn
 &\oplus&
 \bigoplus_{\beta}^{p-|p-b-b'|-1}\R_{2}^\beta\oplus\bigoplus_{\beta}^{b+b'-p-1}\R_{3}^\beta
  \bigg)/\{(1+\delta_{b,0})(1+\delta_{b',0})\}
\label{fullver2}
\eea
where $\R_0^b\equiv0$.
The divisions by $(1+\delta_{b,0})$, for example, ensure that the fusion
rules for $\R_1^0$ match those for $(1,p)$.
Due to (\ref{r11s}), and using
\be
 (r,1)\otimes(r',1)=\bigoplus_{j=|r-r'|+1,\,\mathrm{by}\,2}^{r+r'-1}(j,1)
\ee
the complete set of fusion rules underlying the Kac fusion algebra
is obtained straightforwardly. It can be found in~\cite{Ras1012}.

\subsection{Contragredient Kac fusion rules}
\label{AppContraRules}

Following~\cite{Ras1012}, we introduce
\be
 \Cc_n[(r,s)]=\begin{cases} 
  (r,s),\quad &n>0
  \\[.2cm]
  (r,s)^\ast,\quad &n<0
 \end{cases}
\label{C}
\ee
In our applications, $\Cc_0[(r,s)]$ only appears if $(r,s)$ is fully reducible
in which case
\be
 \Cc_0[(r,s)]=(r,s)=(r,s)^\ast,\qquad s\in\mathbb{Z}_{1,p-1}\cup p\mathbb{N}
\ee
The fusion rules involving contragredient Kac representations are given by or follow readily from
\be
 (r,s)^\ast\otimes(r',s')^\ast=\big((r,s)\otimes(r',s')\big)^\ast,\qquad
 \R_r^b\otimes(r',s')^\ast=\R_r^b\otimes(r',s')
\label{rsrsA}
\ee
and
\bea
 (1,b+kp)\otimes(1,b'+k'p)^\ast&=&\!\!\bigoplus_{j=|k-k'|+2,\,\mathrm{by}\,2}^{k+k'}\!\!\!\!
   \bigoplus_{\beta}^{p-|p-b-b'|-1}\!\!\R_j^\beta\oplus
   \bigoplus_{j=|k-k'|+1,\,\mathrm{by}\,2}^{k+k'-\mathrm{sg}(p-b-b')}\
   \bigoplus_{\beta}^{|p-b-b'|-1}\!\!\R_j^\beta  
\nn
&\oplus&\!\!\bigoplus_\beta^{(b-b')\mathrm{sg}(k'-k)-1}\!\R_{|k-k'|}^\beta
 \oplus\bigoplus_{\beta=|b-b'|+1,\,\mathrm{by}\,2}^{p-|p-b-b'|-1}
 \!\!\Cc_{k-k'}[(1,\beta+|k-k'|p)]
\label{bkbk}
\eea
where $b,b'\in\mathbb{Z}_{0,p-1}$ and $k,k'\in\mathbb{N}_{0}$.
Since $(r,1)$ is irreducible, we thus have
\be
 (r,s)^\ast=(r,1)^\ast\otimes(1,s)^\ast=(r,1)\otimes(1,s)^\ast
\ee
from which it follows that the general fusion product $(r,s)\otimes(r',s')^\ast$ can be
computed as
\be
 (r,s)\otimes(r',s')^\ast=\big((r,1)\otimes(r',1)\big)\otimes\big((1,s)\otimes(1,s')^\ast\big)
\ee
The complete set of fusion rules underlying the contragrediently-extended Kac fusion
algebra can be found in~\cite{Ras1012}.

\section{Dictionary}
\label{AppDic}

Here we present a dictionary for translating the notation used in~\cite{KS0901} 
(and similarly in~\cite{Sut94,FGST0504,FGST0512,FGST0606hep,FGST0606math,NT0902})
for indecomposable quantum-group modules
to the one employed here for $\Wc$-extended modules.
For $s,s'\in\mathbb{Z}_{1,p}$, $a\in\mathbb{Z}_{1,p-1}$ and $n\in\mathbb{N}$, we have
\be
\begin{array}{rll}
 &\Xc_s^+\,\leftrightarrow\,\ketw{1,s},\qquad
 &\Xc_s^-\,\leftrightarrow\,\ketw{2,s}
\\[.3cm]
 &\Pc_a^+\,\leftrightarrow\,\Rh_1^{p-a},\qquad
 &\Pc_a^-\,\leftrightarrow\,\Rh_2^{p-a}
\\[.3cm]
 &\Mc_a^+(n)\,\leftrightarrow\,\ketw{2\cdot n,p-a+(n-1)p},\qquad
 &\Mc_a^-(n)\,\leftrightarrow\,\ketw{1\cdot n,p-a+(n-1)p}
\\[.3cm]
 &\Wc_a^+(n)\,\leftrightarrow\,\ketwa{1\cdot n,a+(n-1)p},\qquad
 &\Wc_a^-(n)\,\leftrightarrow\,\ketwa{2\cdot n,a+(n-1)p}
\end{array}
\ee
where
\be
 \Mc_{p-s}^\mp(1)=\Wc_s^\pm(1)=\Xc_s^\pm
\ee
Direct sums over of the index sets $I_{s,s'}$ and $J_{s,s'}$ in~\cite{KS0901} correspond to
\be
 \bigoplus_{t\in I_{s,s'}}A_t=\bigoplus_{t=|s-s'|+1,\,\mathrm{by}\,2}^{p-|p-s-s'|-1}A_t,\qquad\quad
 \bigoplus_{t\in J_{s,s'}}A_t=\bigoplus_{t}^{s+s'-p-1}A_{p-t}
\ee
that is,
\bea
 \bigoplus_{t\in I_{s,s'}}\Xc_t^{\al}
   &\leftrightarrow&\bigoplus_{j=|s-s'|+1,\,\mathrm{by}\,2}^{p-|p-s-s'|-1}\ketw{\al,j}\nn
 \bigoplus_{t\in I_{s,s'}}\Pc_t^\al
   &\leftrightarrow&\bigoplus_{\beta=|p-s-s'|+1,\, \mathrm{by}\,2}^{p-|s-s'|-1}\Rh_\al^\beta\nn
  \bigoplus_{t\in J_{s,s'}}\Pc_t^\al&\leftrightarrow&\bigoplus_{\beta}^{s+s'-p-1}\Rh_\al^\beta
\eea
Here $\al=+$ ($\al=-$) on the quantum-group side corresponds to
$\al=1$ ($\al=2$) on the logarithmic CFT side.


\end{document}